\documentclass[aps,pra,nofootinbib,twocolumn,floatfix,showpacs,superscriptaddress,10pt]{revtex4-1}
\usepackage[utf8]{inputenc}
\usepackage{amsmath,graphicx,epsfig,amsthm, color, bm}
\usepackage{pifont,bbding,amssymb,soul,mathrsfs}
\usepackage{indentfirst}
\usepackage{times,ragged2e}
\usepackage{appendix, comment} 
\usepackage{CJKutf8}
\usepackage{ucs}
\usepackage[encapsulated]{CJK}
\usepackage{tikz,pgfplots}
\usepackage{booktabs}
\usepackage{hhline}
\usepackage{bbold}
\allowdisplaybreaks

\newcommand{\proj}[1]{\ket{#1}\!\bra{#1}}
\newcommand{\tr}{{\rm tr}}

\newcommand{\Q}{\mathcal{Q}}

\newcommand{\cH}{\mathcal{H}}
\newcommand{\D}{\mathcal{D}}
\newcommand{\bra}[1]{\langle #1|}
\newcommand{\ket}[1]{|#1\rangle}

\newcommand{\mes}{\ket{\Phi^+_d}}

\newcommand{\rhoAB}{\rho_\text{AB}}
\newcommand{\rhoExp}{\rho_{\text{expt}}(v)}
\newcommand{\rhoExpf}{\rho_{\text{expt}, f}(v)}
\newcommand{\rhoW}{W^{(d)}}
\newcommand{\rhoWvd}{\rhoW(v)}
\newcommand{\rhoWv}[1]{W^{(#1)}(v)}
\newcommand{\rhoWf}[1]{W^{(#1)}_f(v)}
\newcommand{\rhoWfd}{W^{(d)}_f(v)}
\newcommand{\Id}{\mathbb{I}}
\newcommand{\SR}{\text{SR}}
\newcommand{\SCHSH}{\mathcal{S}_{\text{\tiny CHSH}}}
\newcommand{\tsz}{t^\star_0}

\definecolor{nblue}{rgb}{0.2,0.2,0.8}

\newtheorem{theorem}{Theorem}
\newtheorem{proposition}[theorem]{Proposition}

\newcommand{\id}{\mathbb{I}}
\newcommand{\Iff}{{\em iff} }

\usepackage[colorlinks=true,citecolor=blue,urlcolor=blue]{hyperref}

\usepackage[capitalize]{cleveref}
\pgfplotsset { compat = 1.14}

\begin{document}
\title{Experimental single-copy distillation of quantumness from higher-dimensional entanglement}

\author{Xiao-Xu Fang}
\affiliation{School of Physics, State Key Laboratory of Crystal Materials, Shandong University, Jinan 250100, China}

\author{Gelo Noel M. Tabia}
\affiliation{Hon Hai (Foxconn) Research Institute, Taipei, Taiwan}
\affiliation{Department of Physics and Center for Quantum Frontiers of Research \& Technology (QFort), National Cheng Kung University, Tainan 701, Taiwan}
\affiliation{Physics Division, National Center for Theoretical Sciences, Taipei 106319, Taiwan}

\author{Kai-Siang Chen}
\affiliation{Department of Physics and Center for Quantum Frontiers of Research \& Technology (QFort), National Cheng Kung University, Tainan 701, Taiwan}

\author{Yeong-Cherng Liang}
\email{ycliang@mail.ncku.edu.tw}
\affiliation{Department of Physics and Center for Quantum Frontiers of Research \& Technology (QFort), National Cheng Kung University, Tainan 701, Taiwan}
\affiliation{Physics Division, National Center for Theoretical Sciences, Taipei 106319, Taiwan}
\affiliation{Perimeter Institute for Theoretical Physics, Waterloo, Ontario, Canada, N2L 2Y5}

\author{He Lu}
\email{luhe@sdu.edu.cn}
\affiliation{School of Physics, State Key Laboratory of Crystal Materials, Shandong University, Jinan 250100, China}
\affiliation{Shenzhen Research Institute of Shandong University, Shenzhen 518057, China}

\date{\today}

\begin{abstract} 
Entanglement is at the heart of quantum theory and is responsible for various quantum-enabling technologies. In practice, during its preparation, storage, and distribution to the intended recipients, this valuable quantum resource may suffer from noisy interactions that reduce its usefulness for the desired information-processing tasks. Conventional schemes of entanglement distillation aim to alleviate this problem by performing collective operations on multiple copies of these decohered states and sacrificing some of them to recover Bell pairs. However, for this scheme to work, the states to be distilled should already contain a large enough fraction of maximally entangled states before these collective operations. Not all entangled quantum states meet this premise. Here, by using the paradigmatic family of two-qutrit Werner states as an exemplifying example, we {\em experimentally} demonstrate how one may use {\em single-copy} local filtering operations to meet this requirement and to recover the {\em quantumness} hidden in these {\em higher-dimensional} states. Among others, our results provide the {\em first} proof-of-principle experimental certification of the Bell-nonlocal properties of these intriguing entangled states, the activation of their usefulness for quantum teleportation, dense coding, and an enhancement of their quantum steerability, and hence usefulness for certain discrimination tasks. Our theoretically established lower bounds on the steering robustness of these states, when they admit a symmetric quasiextension or a bosonic symmetric extension, and when they show hidden dense-codability may also be of independent interest.
\end{abstract}

\maketitle

%\section{Introduction}
{\em Introduction.}--Ever since its discovery, quantum entanglement~\cite{HHHH09} has played a fascinating role in shaping~\cite{Bell64} our understanding of the physical world. In the modern era, entanglement takes the role of a powerful resource, enabling quantum computation~\cite{Jozsa:2003aa,VidalPRL03}, quantum-enhanced communications~\cite{Ekert91,CW92}, and quantum metrology~\cite{Toth:JPA:2014}, etc. To take advantage of these possibilities at a large scale,  a quantum internet~\cite{Wehner:2018aa} where we can readily perform quantum communications between any two points is clearly desirable. Accordingly, the possibility of distributing entanglement in one way or another (see, e.g.,~\cite{ZZH+PRL93,Yin:2017aa}) is surely desirable.

Even with ideal preparation, the quality of entanglement can easily degrade over time, either during the storage or distribution stage. Consequently, much effort has been devoted to understanding how one can recover Bell pairs via entanglement distillation or purification protocols~\cite{Bennett1996PRL,Deutsch1996PRL}. Indeed, for quantum states violating the reduction criterion of separability (RC)~\cite{ReductionCriterion}, iterative single-copy local filtering operations (ScLFs)~\cite{Gisin1996PLA} followed by collective local operations~\cite{Bennett1996PRL,ReductionCriterion} distill, with a nonzero probability, quantum states having an arbitrarily high Bell-state fidelity.

Nonetheless, some bipartite entangled states, e.g., those having positive-partial-transposition~\cite{1996_peres_Phys.Rev.Lett._Separability} (PPT), are  {\em undistillable}~\cite{HHH:PRL:1998}. There is even evidence~\cite{2000_divincenzo_Phys.Rev.A_Evidence,DurPRA00} suggesting that some non-PPT (NPPT) quantum states, such as the {\em Werner states}~\cite{Werner:PRA:1989}, are undistillable. Since all two-qubit entangled states are distillable~\cite{HHH:PRL:1997}, this intriguing phenomenon of bound entanglement~\cite{HHH:PRL:1998} only occurs among {\em higher-dimensional} (HD) mixed entangled states. In contrast, pure HD entanglement may offer an advantage over its qubit counterparts in entanglement distribution~\cite{Steinlechner2017NC,Ecker2019PRX,Hu2020Optica} and various information-processing tasks (see, e.g.,~\cite{LLT+02PRA,LKE+03,Vertesi2010PRL}).

Today, it remains unknown~\cite{HRZPRXQ22} whether all NPPT entangled states are distillable. Even among those distillable, many copies may be required~\cite{Watrous:PRL:2004,FangPRL20,FangPRX22,RegulaPRL22}. Moreover, to distill via the (generalized) recurrence protocol~\cite{Bennett1996PRL,ReductionCriterion}, the input state to the protocol must already have a large enough fully-entangled fraction~\cite{HHH99Teleportation} (to the point that it even ensures its usefulness~\cite{PopescuTB} for teleportation). Thus, the initial noisy entangled state may have to go through additional ScLFs before being subjected to the collective distillation operations. Even though the latter, more technically challenging, operations have been experimentally realized for two copies~\cite{Pan2001Nature,Pan2003Nature,ZYC+03,Yamamoto:2003aa, WaltherPRL05,Reichle:2006aa,Kalb2017Science} (see also~\cite{Hu2021PRLa,Ecker2021PRL}), we are still far from recovering a near-to-perfect Bell pair via conventional distillation schemes.

In contrast, for the sake of recovering quantum advantage over classical resources, ScLFs often suffice. For instance, they can help~\cite{ZGZ+17} distinguish different types of entanglement, recover usefulness for secure communication~\cite{Singh:2021aa}, unveil {\em hidden}~\cite{Pramanik2019PRA,Nery2020PRL,HWL+24,ZFL24}  quantum steerability~\cite{2007_wiseman_Phys.Rev.Lett._Steering,UolaRMP20}, hidden~\cite{LiPRR2021} teleportation power~\cite{PopescuTB} (see also~\cite{Masanes2006PRL}), as well as the (seemingly) hidden Bell-nonlocality of certain two-qubit states~\cite{Gisin1996PLA,Kwiat2001NPhysics,WZHPRL06,WLW+20} and HD states~\cite{PopescuPRL_HNL,Kumari:2024aa} (see also~\cite{Masanes2008PRL,Liang2012PRA}). 
Here, building on the results from~\cite{PopescuPRL_HNL,Liang2012PRA,LiPRR2021}, we experimentally demonstrate the recovery of various quantum features via the {\em same} ScLF for one classic family of HD quantum states---the two-qutrit Werner states. Although tremendous efforts have been devoted to the generation and detection of HD entanglement
and their nonlocal properties (see, e.g.,~\cite{TAZ+PRL04, Huber2010PRL, Schwarz16, Martin2017PRL, Islam2017SA,Bavaresco2018NPhysics,Zeng2018PRL,Wang2018Science, Hu2018SA,Lu2020npjqi,Dada2011Nphysics,Hu2020PRLa,Hu2021PRL,Designolle2021PRL,Qu2022Optica,Qu2022PRL}), our work gives the first experimental demonstration of the {\em single-copy distillation} of quantumness from HD entanglement.

{\em Werner states.}--The bipartite $d$-dimensional ($d \geq 2$) Werner state \cite{Werner:PRA:1989} consists of a convex mixture of the (normalized) projectors $\Pi_{\pm}\equiv \frac{1}{2}\left(\Id_{d^2}\pm V \right)$ onto the symmetric ($+$) and anti-symmetric ($-$) subspace:
\begin{equation}\label{Eq:Werner}
	\rhoWvd:=\frac{v}{n_+^d}\Pi_+^d + \frac{1-v}{n_-^d}\Pi_-^d,\quad v\in \left[ 0,1\right],
\end{equation}
where $\Id_{d^2}$ is the identity operator acting on $\mathbb{C}^d\otimes \mathbb{C}^d$,  $n_\pm^d:=\tr\Pi_\pm^d=\frac{d(d\pm 1)}{2}$ and the swap operator $V$ satisfies $V\ket{\psi}_A\ket{\phi}_B=\ket{\phi}_A\ket{\psi}_B$ for all $\ket{\psi},\ket{\phi}\in\mathbb{C}^d$.
$\rhoWvd$ are PPT~\cite{DurPRA00} and separable~\cite{Werner:PRA:1989} if and only if (\Iff) $v \geq \frac{1}{2}$. However, entangled Werner states are only known to be distillable when~\cite{DurPRA00,2000_divincenzo_Phys.Rev.A_Evidence} $v<v_{\text{distill}}\equiv\frac{d+1}{4d-2}$.  In this regard, a state $\rho$ is said to be distillable~\cite{HH01QIC} if it is $k$-distillable for some integer $k$, i.e., if there exists qubit projectors $\Pi_A,\Pi_B$ such that 
$\frac{\Pi_A\otimes\Pi_B\rho^{\otimes k}\Pi_A\otimes\Pi_B}{\tr(\Pi_A\otimes\Pi_B\rho^{\otimes k}\Pi_A\otimes\Pi_B)}$ is entangled. 
To determine the existence of NPPT bound entangled states~\cite{HRZPRXQ22}, it suffices~\cite{DurPRA00} to consider $\rhoWvd$ for $d>2$.

HD Werner states are not only important because of their role in the NPPT distillability problem but are also peculiar for being useless in various tasks. Indeed, {\em all} HD $\rhoWvd$ satisfy RC~\cite{ReductionCriterion}. Consequently, even those that 
%peculiar, however, not only because of their distillability but also because they are useless in various tasks. Indeed, {\em all} HD $\rhoWvd$ satisfy RC~\cite{ReductionCriterion}. Consequently, even those that 
are distillable cannot be fed directly to the generalized recurrence protocol~\cite{ReductionCriterion} for entanglement distillation. Their compliance with RC also renders them useless for quantum teleportation~\cite{ReductionCriterion} and dense coding~\cite{Bruss2004densecoding}. However, as shown in~\cite{LiPRR2021}, ScLF in the form of qubit projection can activate their usefulness for teleportation whenever $v<v_{\text{distill}}$. 

\begin{figure}[htbp]
	\includegraphics[width=\linewidth]{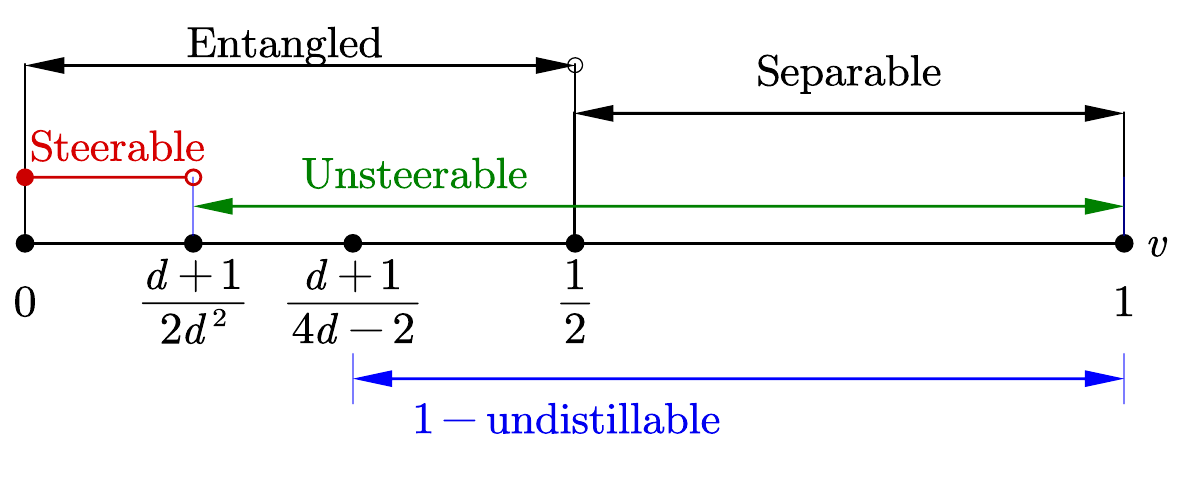}
	\caption{Werner states $\rhoWvd$, $v\in[0,1]$ is entangled \Iff $v < \frac{1}{2}$, 
1-distillable (steerable with projective measurements) when $0 \leq v < v_{\text{distill}} $ ($0 \leq v < v_{\text{steer}} $), where $v_{\text{steer}}=\frac{d+1}{2d^2}<v_{\text{distill}}=\frac{d+1}{4d-2}$. For $d>2$, all $\rhoWvd$ are useless for teleportation, dense-coding, and {\em not} known to violate Bell inequality.}
	\label{Fig1:werner_zone}
\end{figure}

Furthermore, Werner states with $v\ge v_{\text{steer}}\equiv\frac{d+1}{2d^2}$ are the first known examples of entangled quantum states {\em not} violating any Bell~\cite{Werner:PRA:1989} or steering~\cite{2007_wiseman_Phys.Rev.Lett._Steering} inequality with projective measurements (see~\cref{Fig1:werner_zone}). However,  after a successful qubit ScLF~\cite{PopescuPRL_HNL}, $W^{(d\ge 5)}(v_\text{steer})$ does violate the Clauser-Horne-Shimony-Holt (CHSH) Bell inequality~\cite{CHSH}. More generally, $\rhoWvd$ admit~\cite[Theorem 6]{JV23} a symmetric $(1,k)$ or  $(k,1)$-extension~\cite{Terhal03,DPS:PRL:2022}   \Iff $v\ge v_\text{Sym}\equiv\frac{1}{2}\left(1-\frac{d-1}{k}\right)$. In other words, there exists a $(1+k)$-partite qudit state $\rho$ that recovers $\rhoWvd$ after tracing out {\em any} $(k-1)$ copies of Alice's (Bob's) subsystems.
It then follows~\cite{Terhal03} that $\rhoWvd$ for $v\ge v_\text{Sym}$ cannot violate any Bell inequality with $k$  or fewer measurement settings for Alice (Bob)~\cite{Note1}.
In particular, since $v_\text{Sym}$ vanishes when $k=d-1$, $\rhoWvd$ satisfy~\cite{Terhal03} {\em all} Bell inequalities with $d-1$ or fewer measurement bases at one side. For example, all $\rhoWv{3}$ cannot violate any two-or-fewer-setting Bell inequality on one side. Again, a qubit ScLF reveals~\cite{Liang2012PRA}  the Bell-nonlocality (possibly) hidden in some of these states, which we demonstrate experimentally in this work.

How does a qubit ScLF enable the distillation of many of these useful properties? We can gain some intuition into this by first noting that $\Pi_+^d+\Pi_-^d =\Id_{d^2}$ and~\cite{PopescuPRL_HNL}  $\Pi_-^d=\sum_{i<j}\varrho_0^{i,j}$ where
$\varrho_0^{i,j}=\proj{\Psi^- _{i,j}}$ is a singlet projector defined on the two-qubit subspace $\mathcal{H}_{i,j}:=\text{span}\{\ket{k\ell}:=\ket{k}\ket{\ell}\}_{k,\ell=i,j}$. Then, by defining $p:=\frac{1}{d}$, $q:=1-\frac{2d}{d+1}v$, and noting that  $n_-^d$ is the binomial coefficient of $d$ chooses 2, we can rewrite $\rhoWvd$ as a {\em uniform} mixture of two-qubit states $\varrho^{i,j}$: 
\begin{subequations}\label{eq:Wernerstate_q}
\begin{gather}
	\rhoW(q)=\frac{q}{n_-^d}\sum_{i<j}\varrho_0^{i,j}+\frac{1-q}{d^2}\Id_{d^2}=\frac{1}{n_-^d}\sum_{i<j}\varrho^{i,j},\label{Eq:NoisyProjector}\\
	\varrho^{i,j}(q) := q\varrho_{0}^{i,j}+\left( 1-q \right) \left[ p\varrho_{1}^{i,j}+\left( 1-p \right) \varrho_{2}^{i,j} \right],
\end{gather} 
\end{subequations}
where $\varrho_1^{i,j}=\frac{\proj{ii}+\proj{jj}}{2}$, $\varrho_2^{i,j}=\frac{\proj{ij}+\proj{ji}}{2}$.
In this decomposition, only the two-qubit singlet projectors $\{\varrho_0^{i,j}\}_{i,j}$ contribute to the entanglement of $\rhoWvd$. The act of a qubit projection, say, $\Pi_2:=\proj{0}+\proj{1}$, on both sides then keeps the maximally entangled $\varrho_0^{0,1}$ and the separable $\frac{\Id_4}{4}$ with some probability while removing the components in all other subspaces, thereby effectively ``purifying" the entanglement contained in $\rhoWvd$. From~\cref{Eq:NoisyProjector}, we see that upon a successful qubit projection, since $\Pi_2\otimes\Pi_2\rhoWvd\Pi_2\otimes\Pi_2\propto W^{(d=2)}(v')$, we obtain a two-qubit Werner state with a symmetric weight
$v'=\frac{3(d-1)v}{(d+1)(1-v)+3(d-1)v}$. Note that $v'\geq v$ for all $d>2$.

\begin{figure*}[htb]
\includegraphics[width=\linewidth]{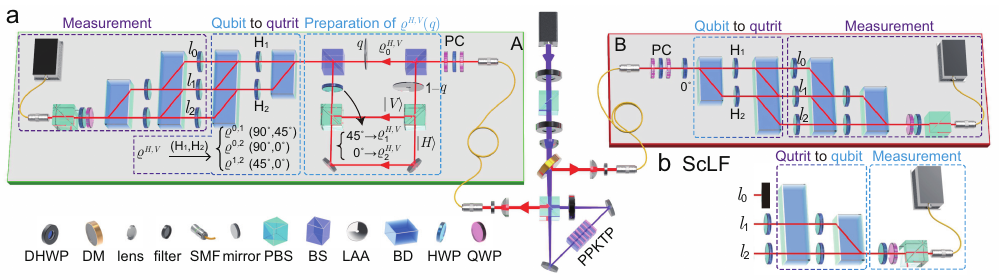} 
\caption{\textbf{a. }Experimental setup to generate $W^{(d=3)}(v)$. We first generate polarization-entangled photons, each coupled into a single-mode fiber (SMF) and sent to Alice and Bob. The fiber-induced polarization drift is corrected by a polarization controller~(PC), which is a half-wave plate~(HWP) sandwiched by two quarter-wave plates~(QWPs). Then, we partially decohere $\varrho^{H,V}_0$ to prepare the state $\varrho^{H,V}(q)$. Finally, $\varrho^{H,V}(q)$ is mapped to $W^{(d=3)}(v)$ encoded in the path DOF with two optical networks, each consisting of two BDs and two HWPs, where the parameter $v$ is adjustable via $q$ by rotating two Linear-Adjustable Attenuators~(LAAs). \textbf{b. }Realization of the ScLF. Optical elements list. DHWP: dual-wavelength HWP@ 405 \& 810~nm. DM: dichroic mirror. NBF: narrow-band filter. PBS: polarizing beam splitter. BS: beam splitter. BD: beam displacer. More details are provided in~\cref{App:Exp}.}

\label{Fig2:SetupWerner}
\end{figure*}

%\section{Experimental Werner state and filtering operation}
{\em Experimental Werner states and filtering operation.}--The decomposition of~\cref{eq:Wernerstate_q} not only sheds light on the relevance of a qubit ScLF but also provides a systematic way for preparing an entangled $\rhoWvd$ for an {\em arbitrary} $d$ 
via generating and mixing the various two-qubit states $\varrho^{i,j}(q)$ residing in $\mathcal{H}_{i,j}$.
To this end, we start by preparing photons pairs maximally entangled in the polarization degree of freedom~(DOF) $\ket{\Psi^+}_\text{AB}=\frac{1}{\sqrt{2}}(\ket{H}\ket{V}+\ket{V}\ket{H})_\text{AB}$, where $H$ and $V$ denote, respectively, horizontal and vertical polarization. As shown in~\cref{Fig2:SetupWerner}a, the entangled photons are generated via the spontaneous parameter down conversion process~(SPDC) from a periodically-poled potassium titanyl phosphate~(PPKTP) crystal in a Sagnac interferometer, which is bidirectionally pumped by a laser with central wavelength at 405~nm. By setting a HWP at 0$^\circ$ on photon B, $\ket{\Psi^+}_\text{AB}$ is transformed into $\varrho_0^{H,V}=\ket{\Psi^-}_\text{AB}\bra{\Psi^-}$. Then, we use two BSs, two PBSs and two tunable LAAs~(with transmittance of $q$ and $1-q$, respectively) to partially decohere $\varrho_0^{H,V}$ and prepare the mixed state $\varrho^{H,V}(q)$ in~\cref{eq:Wernerstate_q} by randomly rotating a HWP between $45^\circ$ and $0^\circ$. 
Finally, we obtain the $d$-dimensional Werner states $\rhoWvd$ from $\varrho^{H,V}(q)$ with a qubit-to-qudit mapping, where the qudit is encoded in the path DOF $\{\ket{l_k}\}$ and the mapping is achieved with two optical networks each consisting of $d-1$ beam displacers (BDs) and a series of HWPs on photon A and B. For $d=3$ in this work, we use two BDs and two HWPs at each side~(as shown in~\cref{Fig2:SetupWerner}a) to prepare eleven $W^{(d=3)}(v)$ with $v\in[0,0.5]$ in steps of 0.05. See~\cref{App:WernerPrep} for further details.

\begin{figure*}[htbp]
	\includegraphics[width=\linewidth]{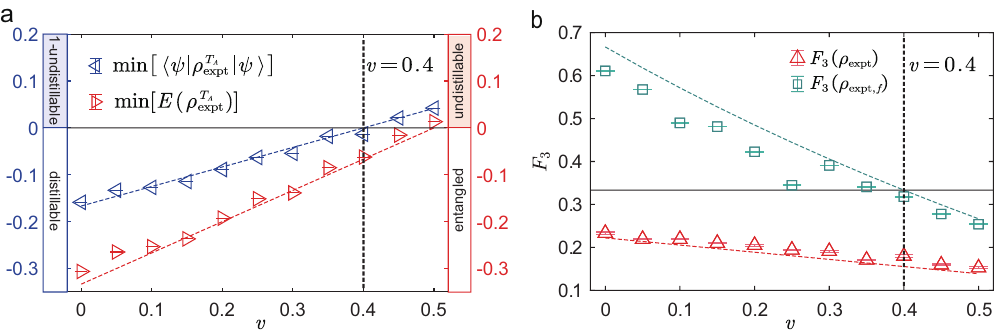} 
	\caption{\textbf{a. } Experimental results confirming the entanglement and 1-distillability of the experimentally prepared Werner states $\rhoExp$.  The red right triangle $\vartriangleright$ gives the minimum eigenvalue of $\rho_{\text{expt}}^{\text{T}_A}$, where a negative (non-negative) value confirms entanglement (undistillability). Similarly, a negative (non-negative) value for the blue triangle $\vartriangleright$ confirms distillability (1-undistillability). See text for details. The vertical dashed line marks the 1-distillability boundary. \textbf{b. }~Theoretical (dashed lines) and experimental (dots) results illustrating the teleportation power before and after applying the qubit ScLF on $\rhoExp$. The bottom (red) results correspond to the unfiltered states $W^{(d=3)}(v)$ and $\rhoExp$ (experiment), while the top (green square) results are for the filtered state $W^{(d=3)}_f(v)$ and $\rhoExpf$ (experiment). The solid black line is the classical threshold $F^c_{d=3}=\frac{1}{3}$ and the vertical dashed line marks the boundary of usefulness for teleportation.
}
	\label{Fig3:distill_telep}
\end{figure*}

We reconstruct the prepared $\rhoExp$ using quantum state tomography, in which an iterative maximum-likelihood algorithm~\cite{2005_altepeter_AdvancesInAtomicMolecularandOpticalPhysics_Photonic} is performed on the twofold coincidence collected in $9 \times 9$ measurement basis vectors (see~\cref{App:Meas}).  We then calculate the reconstructed $\rhoExp$'s fidelity for $W^{(d=3)}(v)$, and find that they fall within $[0.958, 0.995]$ (see \cref{App:Fidelity} for details). Also, we  confirm the entanglement of $\rhoExp$ for $v\le0.45$ from their NPPT property (\cref{Fig3:distill_telep}, $\vartriangleright$) and their 1-distillability~\cite{DurPRA00} for $v\le 0.4$ by showing that, upon optimizing over entangled two-qubit state $\ket{\psi}$, $\text{min}_{\ket{\psi}}\left[\bra{\psi}\rho_{\text{expt}}^{\text{T}_{A}}(v)\ket{\psi}\right]<0$ (see \cref{Fig3:distill_telep}, $\vartriangleleft$); here, $(.)^{\text{T}_{A}}$ is the partial transposition with respect to $A$. Furthermore, we verify the separability of $\rho_{\text{expt}}(v=0.5)$ by recalling from~\cite[Corollary 3]{2002_gurvits_Phys.Rev.A_Largest} that $\rho$ acting on $\mathbb{C}^d\otimes\mathbb{C}^d$ is separable if it satisfies $\|\rho-\frac{\Id_{d^2}}{d^2}\|_{2}^{2} \leqslant \frac{1}{d^2(d^2-1)}$. These results agree well with the theoretical predictions except that $\rho_{\text{expt}}(v=0.4)$ appears to be 1-distillable while $W^{(d=3)}(v=0.4)$ is not. 

{\em ScLF and distillation of quantumness.}--With our encoding of the quantum information, a qubit filtering---relevant for distilling the various quantum features alluded to above---is straightforward. To keep only contributions to the two lower optical paths $\{\ket{l_1},\ket{l_2}\}$, we block the upper optical path $\ket{l_0}$ (\cref{Fig2:SetupWerner}b). After ScLF, the two-qubit state $\rhoExpf$ encoded in $\{\ket{l_1},\ket{l_2}\}$ is mapped back to the polarization DOF and reconstructed using a HWP, a QWP and a PBS---their fidelity ranges roughly from 0.934 to 0.995. In the following, we compare the properties of $\rhoExp$ and $\rhoExpf$ to illustrate how ScLF has enabled the recovery of quantum features missing in $\rhoExp$.

We begin with their Bell-nonlocality property. Except for $\rho_{\text{expt}}(v=0)$, which is {\em not} even 2-quasiextendible, all other $\rhoExp$ are symmetric $(1,k)$ or $(k,1)$ quasi-extendible \Iff the corresponding $\rhoWvd$ are. Hence, $\rhoExp$ with $v\ge \frac{1}{2}(1-\frac{2}{k})\neq 0$ cannot violate any Bell inequalities with $k$ or fewer measurement bases at one side. In fact, despite extensive numerical optimizations, we do not find any Bell-inequality violation by $\rho_{\text{expt}}(v)$; see~\cref{App:BellNonlocality}. In contrast, using the Horodecki criterion~\cite{1995_horodecki_PhysicsLettersA_Violating}, the filtered states $\rhoExpf$ give a maximal CHSH-Bell value~\cite{CHSH} of $\SCHSH=2.65\pm 0.02, 2.46\pm 0.01, 2.02\pm0.02$, and $2.01\pm0.01$ for $v= 0, 0.05, 0.10$, and $0.15$, respectively, in agreement with the results derived for $\rhoWvd$~\cite{Liang2012PRA}. Note, however, that $\rho_{\text{expt}, f}(v=0.1)$ and $\rho_{\text{expt}, f}(v=0.15)$ violates the CHSH-Bell inequality ($\SCHSH\le 2$) by less than one standard deviation. 

Next, recall from~\cite{HHH99Teleportation} that a quantum state $\rho$ is useful for teleporting an unknown two-qudit state if its fully-entangled fraction: 
\begin{equation}\label{eq:FEF}
	F_{d}(\rho)=\max_{\ket{\Psi_d}} \bra{\Psi_d}\rho\ket{\Psi_d},
\end{equation}
is larger than $F^c_{d}=\frac{1}{d}$, where $\ket{\Psi_d} = (\id_d \otimes U_d) \mes$ is an arbitrary {\em maximally entangled} two-qudit state, $\mes =\frac{1}{\sqrt{d}}\sum_{i=0}^{d-1}\ket{i}\ket{i}$, while $\id_d$ and $U_d$ are the identity and an arbitrary unitary matrix acting on $\mathbb{C}^d$, respectively. From~\cref{eq:FEF}, one can easily verify that if a {\em two-qubit} state $\rho$ satisfies $F_2>\frac{1}{2}$, it must also satisfy $F_d>\frac{1}{d}$ for all integer $d>2$ (see \cref{App:FEF}).
Since all two-qubit states violating the CHSH-Bell inequality satisfy~\cite{Horodecki:1996aa} $F_2>\frac{1}{2}$, the CHSH-Bell violation of $\rhoExpf$ for $v\le  0.05$ already confirms, modulo statistical uncertainty and the independent and identical trials {\em assumption}, their usefulness for teleportation. In fact, by using \cref{eq:FEF} and the explicit form of $\rhoExpf$, we see from~\cref{Fig3:distill_telep}b that the teleportation power of all $\rhoExp$ with $v<v_\text{distill}$ is successfully recovered via our qubit ScLF, in agreement with the results of~\cite{LiPRR2021}. In~\cref{App:DenseCodability}, we show that the same ScLF also activates the dense-codability~\cite{Bruss2004densecoding} of some $\rhoWvd$ and provide experimental evidence in~\cref{App:ExpDenseCodability} confirming these findings.

\begin{figure}[h!tbp]
	\includegraphics[width=\linewidth]{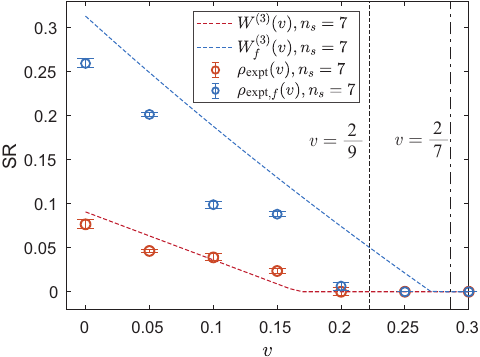}
	\caption{
	Best lower bound on the steering robustness $\SR$ found for $W^{(3)}(v)$, $W^{(3)}_f(v)$ [the state obtained from $W^{(3)}(v)$ via qubit ScLF], and their experimental counterparts $\rhoExp$ and $\rhoExpf$. The vertical dashed (dashed-dotted) line at $v=\frac{2}{9}$ ($v=\frac{2}{7}$) marks the boundary of unsteerability with projective measurements before (after) filtering. See~\cref{App:SR,App:ExpSR} for details. 
}
	\label{Fig4:SR_th_exp}
\end{figure}

Finally, we compare the steering robustness~\cite{UolaRMP20} $\SR$---a quantifier of steerability monotonously related~\cite{PianiPRL15} to the usefulness of a state for some channel discrimination task---between $\rhoExp$ and $\rhoExpf$. To our knowledge, the exact value of $\SR[\rhoWvd]$ is unknown. In~\cref{App:SR}, we provide the best lower bound on $\SR[\rhoWvd]$ we have found~\cite{2017_cavalcanti_Rep.Prog.Phys._Quantum} by considering up to $n_s=7$ measurement bases. Accordingly, we see in~\cref{Fig4:SR_th_exp} an improvement from these lower bounds on $\SR[\rhoExp]$ to $\SR[\rhoExpf]$ for all $v<0.25$. In particular, the steerability of $\rhoExpf$ for $v=0.2$ is clearly witnessed while that of $\rhoExp$ is unknown. Note further from~\cref{Fig4:SR_th_exp} that, in principle, steerability may also be activated beyond $v>v_\text{steer}=\frac{2}{9}$ for $d=3$. However, due to experimental imperfections, we do not observe the activation for $v=0.25$.

%\section{Conclusion}
{\em Conclusion.}--Using local {\em qubit} filtering operations, we experimentally demonstrate the single-copy distillation of several {\em quantum features} from the paradigmatic family of higher-dimensional (two-qutrit) Werner states. Via the coordinated local operations of $U\otimes U$-twirling, one can always transform {\em any} bipartite quantum state to a Werner state.  Thus, our experimental demonstration of single-copy distillation also applies to {\em all} quantum states that transform to a 1-distillable two-qutrit Werner state by these local operations. In turn, this also demonstrates a key step involved in the conventional entanglement distillation~\cite{ReductionCriterion} from higher-dimensional quantum states via Werner states.

To experimentally prepare Werner states, we provide an intuitive decomposition of any finite-dimensional Werner states as a mixture of two-qubit states (see also~\cref{App:WernerDecomposition-Any}). The decomposition, in turn, sheds light on the relevance of qubit filtering operations for these states, partially addressing one of the open problems from~\cite{LiPRR2021}. Conceivably, our experimental scheme for converting qubit states to higher-dimensional states may also find applications in preparing other higher-dimensional states. 

Among the various quantum features distilled from the experimentally prepared two-qutrit Werner states, their Bell nonlocality is especially noteworthy. Indeed, our work provides the first proof-of-principle validation of the Bell nonlocality of these important higher-dimensional states, {\em hidden} or otherwise. Despite experimental imperfections, we have also recovered the missing teleportation power in some of these higher-dimensional states.  Using the lower bound on steering robustness we have established theoretically (see~\cref{App:SR}), our experimental results also provide strong evidence for steering robustness enhancement  (and hence usefulness for certain discrimination tasks~\cite{PianiPRL15}) via single-copy distillation.  The former theoretical results may also be of independent interest, e.g., in the studies of measurement incompatibility~\cite{SLChen16}.

Note again that, in contrast with other entanglement distillation protocols~\cite{Bennett1996PRL,Deutsch1996PRL,BBPS96,ReductionCriterion}, our ScLFs only require local manipulations at {\em individual} copies of the noisy states, thus making them practical, especially in a photonic setup. Indeed, given its simplicity and effectiveness, it is surely worth exploring their implementation also in other quantum information processing platforms, e.g., solid state systems. For future work, exploring the power of ScLFs beyond a qubit projection is clearly of interest. For example, could a higher-dimensional local filtering operation be more effective, e.g., in demonstrating the hidden nonlocality of more $\rhoWvd$ or other states? If they exist, we can always experimentally implement them (see ~\cref{App:ScLFS_Beyond_Qubit}), transforming a $d$-dimensional state to a $d'$-dimensional state (for {\em any} $d'\le d$). Similarly, if such operations for activating dense codability exist, the advantage gained in implementing them might outweigh the need for classical communication involved in their realization, thereby overcoming the objection raised in~\cite{Bruss2004densecoding} for allowing such operations in the characterization of dense-codability. However, since Werner states admit the qubit decomposition of \cref{eq:Wernerstate_q}, we conjecture that a more effective HD filter does not exist. An analytic proof confirming our numerical observation of when a Werner state admits a quasi-symmetric extension or a bosonic symmetric extension is desirable, too. 

\begin{acknowledgments}
{\em Acknowledgments.}--
We thank Andrew Doherty, Huan-Yu Ku, Paul Skrzypczyk, and Yujie Zhang for useful discussions. KSC thanks the hospitality of the Institut N\'eel, where part of this work was completed. This work is supported by the National Key R\&D Program of China~(Grants No. 2019YFA0308200), the Shandong Provincial Natural Science Foundation~(Grant No. ZR2023LLZ005), the Taishan Scholar of Shandong Province~(Grants No. tsqn202103013), the Shenzhen Fundamental Research Program (Grant
No. JCYJ20220530141013029), the National Science and Technology Council, Taiwan (Grants No. 109-2112-M-006-010-MY3, 112-2628-M-006-007-MY4, 113-2917-I-006-023, 113-2918-I-006-001) and the Foxconn Research Institute, Taipei, Taiwan. This research was supported in part by the Perimeter Institute for Theoretical Physics. Research at Perimeter Institute is supported by the Government of Canada through the Department of Innovation, Science, and Economic Development, and by the Province of Ontario through the Ministry of Colleges and Universities.
\end{acknowledgments}

\appendix

\section{Experimental details}\label{App:Exp}

Here, we provide further descriptions of our experimental setup. 
In our experiment, the Jones matrix $U_\text{\tiny HWP}$ $(U_\text{\tiny QWP})$ of the half (quarter)-wave plate is a unitary:
\begin{equation}
\begin{aligned}
U_\text{\tiny HWP}(\theta)&=-\left(\begin{array}{cc}
\cos 2 \theta & \sin 2 \theta \\
\sin 2 \theta & -\cos 2 \theta
\end{array}\right), \\ U_\text{\tiny QWP}(\theta)&=\frac{1}{\sqrt{2}}\left(\begin{array}{cc}
1+i \cos 2 \theta & i \sin 2 \theta \\
i \sin 2 \theta & 1-i \cos 2 \theta
\end{array}\right).
\end{aligned}
\end{equation}
where $\theta$ is the angle between the fast axis and the vertical direction. The global phase in front of $U_\text{\tiny HWP}$ can be ignored because a HWP is present in each path of the experimental setups. To follow the subsequent discussions, we remind the readers that a PBS (BD) transmits a photon with horizontal (vertical) polarization but reflects (deviates) one with vertical (horizontal) polarization.

%\emph{\textbf{Entanglement source}.---}
\subsection{Entanglement source}
\begin{figure}[htbp]
\includegraphics[width=0.9\columnwidth]{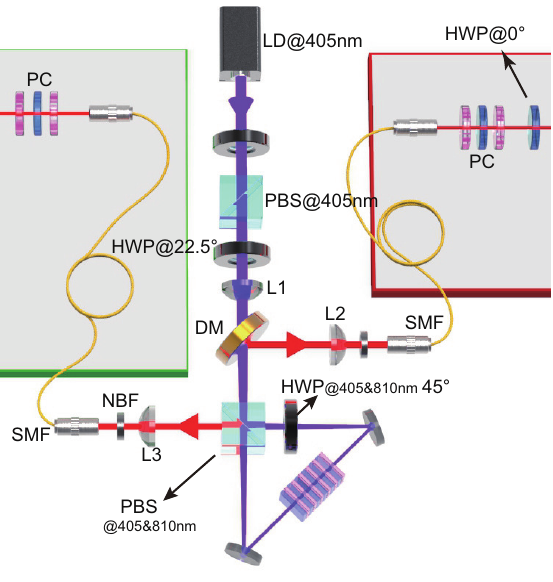} 
\caption{Zoom-in view of the ``Entanglement source" part of the experimental setup shown in Fig.~2a in main text for generating maximally entangled photon pairs.}
\label{Fig:Setup_pent}
\end{figure}
This part, as shown in~\cref{Fig:Setup_pent}, aims to generate a maximally entangled state in the form of $\ket{\Psi^{+}}_\text{AB}=\frac{1}{\sqrt{2}}(\ket{H}\ket{V}+\ket{V}\ket{H})_\text{AB}$.  
The polarization of $405$~nm pump light from a laser diode~(LD) is set at $\ket{H_p}$ with a HWP and a PBS, where the HWP is used to adjust the power intensity of pump light. Then, we use the second HWP set at 22.5$^\circ$, $\ket{H_p}$ to rotate polarization, i.e., $\ket{+_p}=\frac{1}{\sqrt{2}}(\ket{H_p}+\ket{V_p})$. 
Subsequently, the pump beam is focused into a PPKTP crystal with a beam waist of 74 $\mu$m by two lenses L$_{1}$ with focal lengths of $75$~mm and $125$~mm, respectively. The PPKTP crystal with a poling period of $\Lambda=10.025~\mu$m is held in a copper oven with its temperature controlled by a temperature controller set at $29^{\circ}$C to realize the optimum type-\uppercase\expandafter{\romannumeral2} quasi-phase matching at $810$~nm. 

After passing through a dichroic mirror, the pump beam arrives at a dual-wavelength PBS, where it is converged by lenses and split by the PBS to pump the PPKTP crystal coherently in both the clockwise and counterclockwise directions. 
The clockwise- and counterclockwise-propagating photons are then recombined by the dual-wavelength PBS to generate polarization-entangled photon pairs with the ideal form of $\ket{\Psi^{+}}_\text{AB}=\frac{1}{\sqrt{2}}(\ket{H}\ket{V}+\ket{V}\ket{H})_\text{AB}$. 
These photons are then filtered by a narrow band filter~(NBF) with a full width at half maximum~(FWHM) of $3$~nm and each coupled into a single-mode fiber~(SMF) by lenses with focal length of $200$~mm (L$_2$ and L$_3$) and objective lenses. 

For our experiment, the power intensity of pump light is set at $5$~mW, and we observe a twofold coincidence count rate of 73~kHz. The two photons are coupled out by two collimators, and the polarization of each photon is corrected by polarization controller~(PC) consisting two QWPs and a HWP. Finally,  we let photon B go through an additional HWP at 0$^{\circ}$ to produce the singlet state $\varrho_0^{H,V}=(\proj{\Psi^{-}})_\text{AB}$, where $\ket{\Psi^{-}}_\text{AB} = \frac{1}{\sqrt{2}}\ket{H}\ket{V}-\ket{V}\ket{H})_\text{AB}$.

\subsection{Two-qutrit Werner state preparation}
\label{App:WernerPrep}

\begin{figure}[h!tb]
\includegraphics[scale=0.9]{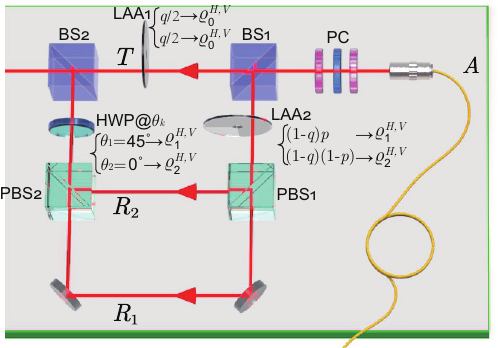}
\caption{Experimental setup that generates the mixed state of \cref{eq:rho_a2b2}, where the different components are distinguishable by their temporal DOF. To generate $p\varrho_1^{H,V}+(1-p)\varrho_2^{H,V}$, we randomly set a HWP at 45$^\circ$ or 0$^\circ$ and attenuate the beam accordingly by changing the LAA$_2$ setting.}
\label{Fig:Setup_WernerPre1}
\end{figure}

Using the singlet state $\varrho_0^{H,V}$, we can further prepare the state $\varrho_2^{H,V}=\frac{\ket{HV}\!\bra{HV}+\ket{VH}\!\bra{VH}}{2}$ by decohering $\varrho_0^{H,V}$ in the $\{\ket{H},\ket{V}\}$ basis, and obtaining therefrom the state $\varrho_1^{H,V}=\frac{\proj{HH}+\proj{VV}}{2}$ via the bit-flip operation $\ket{H}\leftrightarrow\ket{V}$ on one side.
Indeed, photons traveling through different paths $R_1$, $R_2$, and $T$ (see~\cref{Fig:Setup_WernerPre1}) form an incoherent mixture because the time difference between them exceeds their coherence length. Specifically, the two mixed states $\varrho_1^{H,V}$ and $\varrho_2^{H,V}$ can be generated when the HWP after PBS$_2$ is set at $\theta_1=45^{\circ}$ and $\theta_2=0^{\circ}$, respectively. The step-by-step calculation leading from $\varrho_0^{H,V}$ to the two $\varrho_k^{H,V}$'s is as follows:
\begin{equation}\label{eq:WernerPre1}
	\begin{split}
	\ket{\Psi^-}_\text{AB}&=\frac{1}{\sqrt{2}}(\ket{HV}-\ket{VH})_\text{AB}\\
	&\xrightarrow[\text{photon A}]{\text{PBS}_1}\frac{1}{\sqrt{2}}(\ket{HV} \ket{R_1}-\ket{VH}\ket{R_2})_\text{AB}\\
	&\xrightarrow[\text{photon A}]{\text{PBS}_2}\frac{1}{2}(\proj{HV}+\proj{VH})_\text{AB}\\	
    	&\xrightarrow[\text{photon A}]{\text{HWP}~@~\theta_k(t_k)}\varrho_k^{H,V}.
	\end{split}
\end{equation}

After mixing the states prepared through different path lengths and at different times, we obtain the desired polarization-encoded two-qubit mixed states
\begin{equation}\label{eq:rho_a2b2}
	\varrho^{H,V}(q)=q\varrho_0^{H,V}+(1-q)\left[p\varrho_1^{H,V}+(1-p)\varrho_2^{H,V}\right],
\end{equation}
where the mixing parameters $q=1-\frac{2d}{d+1}v$ and $p=\frac{1}{d}$ are achieved by adjusting the LAA$_1$ and LAA$_2$ according to the transmittance of the tested light through the different paths.

\begin{figure*}[h!tb]
\includegraphics[scale=1]{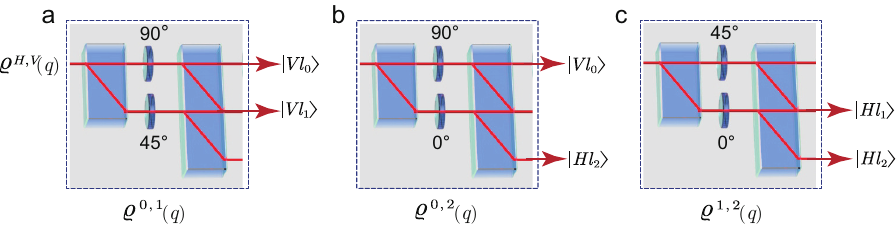}
\caption{Experimental setups to introduce the path DOF and produce $\rho_{\text{expt}}$, which is a uniform mixture of $\varrho^{0,1}$, $\varrho^{0,2}$, and $\varrho^{1,2}$. The different configurations of HWP in a, b, and c result in photons exiting from the BD via different paths. We encode the current paths with the corresponding polarization $\ket{Vl_0}$, $\ket{Hl_1}$ or $\ket{Vl_1}$ and $\ket{Hl_2}$, respectively, as $\ket{0}$, $\ket{1}$ and $\ket{2}$.}
\label{Fig:Setup_wernerBD2}
\end{figure*}

Via an appropriate combination of BDs and HWPs, we can convert the polarization-encoded $\varrho^{H,V}(q)$ to a two-qubit state $\varrho^{i,j}(q)$ encoded in the $i$-th and $j$-th path DOF. Mixing them uniformly then gives the desired $d$-dimensional Werner state:
\begin{equation}
    W^{(d)}(q)=\frac{2}{d(d-1)}\sum_{i<j} \varrho^{i,j}(q).
\label{eq:Wernerstate_a3b3}
\end{equation}
Consider, e.g., the preparation of the two-qutrit Werner states shown in \cref{Fig:Setup_wernerBD2}. We adjust the HWPs at both A and B to inject $\varrho^{H, V}$ into $\binom{3}{2}$ different paths at each time. Then, mixing uniformly the states $\varrho^{0,1}$, $\varrho^{0,2}$, and $\varrho^{1,2}$ prepared at different times give an ensemble corresponding to \cref{eq:Wernerstate_a3b3} for $d=3$. The detailed preparation process is shown in \cref{Fig:Setup_wernerBD2}. Specifically, using the HWP settings of \cref{Fig:Setup_wernerBD2}a, \cref{Fig:Setup_wernerBD2}b, and \cref{Fig:Setup_wernerBD2}c, respectively, the horizontally (vertically) polarized photons will exit from the last BD via path $l_1 (l_0)$, $l_2 (l_0)$, and $l_2 (l_1)$. Together, their average gives the desired $\rho_{\text{expt}}$ encoded in the spatial modes $\{\ket{l_k}\}_{k=0,1,2}$.

\twocolumngrid

\subsection{Measurement of the two-qutrit Werner states}
\label{App:Meas}

Note that by construction and the nature of BD, in every single run, the local photonic quantum states that appear in each path $l_k$ always come with a {\em definite} polarization:
\begin{equation}\label{Eq:DOFs}
	l_0 \leftrightarrow \ket{V},\quad
	l_1 \leftrightarrow \ket{V} \text{ or } \ket{H},\quad
	l_2 \leftrightarrow \ket{H}.
\end{equation}
Even when the photon exits via $l_1$, we know by construction whether the corresponding photonic polarization state is $\ket{V}$ or $\ket{H}$. Since the polarization DOF is decoupled from the spatial DOF and each term in the sum in \cref{eq:Wernerstate_a3b3}  involves only two of the spatial modes (see~\cref{Fig:Setup_wernerBD2}), we can rely on polarization measurements to realize any qutrit measurement relevant to our experiment.

\begin{figure}[h!tb]
\includegraphics[width=1\columnwidth]{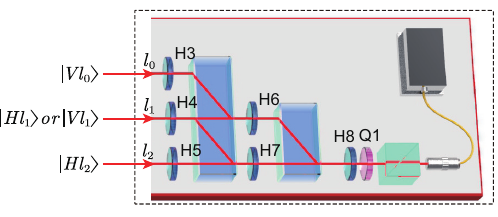}
\caption{Experimental setups that measure the path information encoded in our qutrit photon contain six HWPs, two BDs, one QWP, a PBS, and a detector. Corresponding to the preparation of the state, we mark the optical paths during the measurement as $l_0$, $l_1$, and $l_2$.}
\label{Fig:Setup_qutritM}
\end{figure}

To this end, the path-encoded qutrit state is analyzed through a measurement device consisting of six HWPs, two BDs, one QWP, and one PBS (see \cref{Fig:Setup_qutritM}). For QST, we perform an overcomplete set of path measurements with nine basis vectors via the correspondence of~\cref{Eq:DOFs}:
\begin{equation}\label{Eq:Ms}
\begin{gathered}
	\text{M1}:~\ket{0},\quad \text{M2}:~\ket{1},\quad \text{M3}:~\ket{2},\\
	\text{M4}:~1/\sqrt{2}(\ket{0}+\ket{1}),\quad \text{M5}:~1/\sqrt{2}(\ket{0}+i\ket{1}),\\
	\text{M6}:~1/\sqrt{2}(\ket{0}+\ket{2}),\quad \text{M7}:~1/\sqrt{2}(\ket{0}+i\ket{2}),\\
	\text{M8}:~1/\sqrt{2}(\ket{1}+\ket{2}),\quad \text{M9}:~1/\sqrt{2}(\ket{1}+i\ket{2}).
\end{gathered}
\end{equation}
In particular, the wave plate settings needed to perform a projection onto the individual states of~\cref{Eq:Ms} are shown in~\cref{tab:M}.
Note that the HWPs H3, H4, and H5 remove the original polarization information and measure the path information at the same time. Therefore, the HWP H4 should be set according to the polarization of the photon traveling through path $l_1$ while the HWP H3 and H5 may be set, without loss of generality, at $45^{\circ}$. For example, if we want to measure M2, H4 should be set to $0^{\circ}$ for $\ket{Hl_1}$ and $45^{\circ}$ for $\ket{Vl_1}$. 
Then, the light will pass through the HWP H7 in both cases. From the collection of $9\times9$  twofold coincidence counts, we reconstruct our density matrix using an iterative maximum-likelihood algorithm~\cite{2005_altepeter_AdvancesInAtomicMolecularandOpticalPhysics_Photonic}.

\begin{table}[htbp]
\centering
\caption{Waveplates settings in \cref{Fig:Setup_qutritM} to achieve a single qutrit measurement. $\theta$ means that the plate can be set at any angle, but for simplicity, we can always set HWP at $45^{\circ}$. The first (second) entry in the third column is the setting} for a $\ket{H}$ ($\ket{V}$) photon traveling through $l_1$. 
\begin{tabular}{|c|c|c|c|c|c|c|c|}
    \hline
    Basis vector    &H3    & H4 ($H$; $V$)   & H5   & H6   & H7   & H8  & Q1   \\ \hline
    M1  &  $45^{\circ}$    & $\theta$   & $\theta$   & $0^{\circ}$   & $\theta$ & $0^{\circ}$   & $0^{\circ}$   \\ \hline
    M2 &$\theta$    & $0^{\circ} \text{; } 45^{\circ}$    & $\theta$   & $\theta$   & $45^{\circ}$   & $45^{\circ}$   & $0^{\circ}$   \\ \hline
    M3 &$\theta$    & $\theta$   & $45^{\circ}$   & $\theta$   & $90^{\circ}$  & $45^{\circ}$   & $0^{\circ}$   \\ \hline
    M4    &$45^{\circ}$    & $0^{\circ} \text{; } 45^{\circ}$    & $\theta$   & $0^{\circ}$   & $45^{\circ}$ & $22.5^{\circ}$   & $0^{\circ}$   \\ \hline
    M5   &$45^{\circ}$    & $0^{\circ} \text{;  } 45^{\circ}$    & $\theta$   & $0^{\circ}$   & $45^{\circ}$ & $0^{\circ}$   & $45^{\circ}$   \\ \hline
    M6   &$45^{\circ}$    & $\theta$    & $45^{\circ}$   & $0^{\circ}$   & $90^{\circ}$ & $22.5^{\circ}$  & $0^{\circ}$   \\ \hline
    M7    &$45^{\circ}$    & $\theta$    & $45^{\circ}$   & $0^{\circ}$   & $90^{\circ}$ & $0^{\circ}$   & $45^{\circ}$   \\ \hline
    M8   &$\theta$    & $45^{\circ} \text{; } 90^{\circ}$    & $45^{\circ}$   &$45^{\circ}$   & $90^{\circ}$ & $22.5^{\circ}$   & $0^{\circ}$   \\ \hline
    M9  &$\theta$    & $45^{\circ} \text{; } 90^{\circ}$    & $45^{\circ}$   & $45^{\circ}$   & $90^{\circ}$ & $0^{\circ}$  & $45^{\circ}$   \\ \hline         
\end{tabular}%
\label{tab:M}%
\end{table}%

\subsection{Filtering operation and two-qubit QST}
\label{App:QubitQST}

\begin{figure}[h!tb]
\includegraphics[width=1\columnwidth]{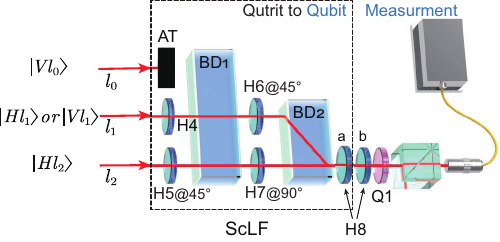}
\caption{Experimental setups for performing the local filtering operation and the measurements of the filtered two-qubit states. The photons traveling through the upper optical path $\ket{l_0}$ are blocked via an absorbing termination~(AT).}
\label{Fig:Setup_filter}
\end{figure}

As mentioned above, our two-qutrit Werner state is encoded in the spatial DOF, decoupled from its polarization DOF,
so the photonic polarization state could be changed at will by changing the setting of the HWPs H4 at both sides A and B. 
For definiteness, we make all the photons vertically polarized before the $\text{BD}_1$ in \cref{Fig:Setup_filter}.

After the local filtering operation, the states change from a qutrit to a qubit, and we convert the encoding to polarization.
With respect to \cref{Fig:Setup_filter}, we see that for an arbitrary qutrit state 
\begin{equation}
	\ket{\psi(\alpha,\beta,\gamma)}:= \frac{\alpha \ket{0}+\beta \ket{1}+\gamma \ket{2}}{\sqrt{\alpha^2+\beta^2 + \gamma^2}}
\end{equation}	
encoded in the spatial DOF of our setup, BD$_1$ and the HWP H6 transform $\ket{\psi(\alpha,\beta,\gamma)}$ as follows:
\begin{equation}\label{eq:filter_stepbystep}
    \begin{aligned}
    &\ket{\psi(\alpha,\beta,\gamma)}\sim\frac{\alpha \ket{Vl_0}+\beta \ket{H \text{ or } V}\ket{l_1}+\gamma \ket{Hl_2}}{\sqrt{\alpha^2+\beta^2+\gamma^2}}
    \\
    &\xrightarrow[\text{photon on path}~l_0]{\text{block}}\frac{\beta \ket{H \text{ or } V}\ket{l_1}+\gamma \ket{Hl_2}}{\sqrt{\beta^2+\gamma^2}}\\   
    &\xrightarrow[\text{photon on path}~l_1]{\text{H4}}\frac{\beta \ket{Vl_1}+\gamma \ket{Hl_2}}{\sqrt{\beta^2+\gamma^2}}\\   
    &\xrightarrow[\text{photon on path}~l_2]{\text{H5}@45^{\circ}}\frac{\beta \ket{Vl_1}+\gamma \ket{Vl_2}}{\sqrt{\beta^2+\gamma^2}}\\   
    &\xrightarrow[\text{photon on path}~l_1,l_2]{\text{BD}_1}\frac{\beta \ket{Vl_1}+\gamma \ket{Vl_2}}{\sqrt{\beta^2+\gamma^2}}\\
    &\xrightarrow[\text{H7}@90^{\circ} \text{on }l_2]{\text{H6}@45^{\circ} \text{on } l_1}\frac{\beta \ket{Hl_1}+\gamma \ket{Vl_2}}{\sqrt{\beta^2+\gamma^2}}\\
    &\xrightarrow[\text{photon on path}~l_1,l_2]{\text{BD}_2}\frac{\left(\beta \ket{H}+\gamma \ket{V}\right)\ket{l_2}}{\sqrt{\beta^2+\gamma^2}}\sim\ket{\psi(\beta,\gamma)}\\    
    \end{aligned}
\end{equation}

By encoding $\ket{H}\to\ket{1}$ and $\ket{V}\to\ket{2}$, we obtain the qubit state
\begin{equation}
	\ket{\psi(\beta,\gamma)}:= \frac{\beta \ket{1}+\gamma \ket{2}}{\sqrt{\beta^2+ 
    \gamma^2}},
\end{equation}	
which means that the qubit filtering operation is performed successfully. Note that these local operations already suffice for our experimental demonstration of single-copy distillation of quantum features. However, in the experiment, we have followed \cite{LiPRR2021} to apply, additionally, the qubit rotations $\sigma_x$ and $\sigma_z$, see~\cref{Eq:ActualFilteredState}, thereby leading to different local filtering operations on the two sides. The qubit rotations can be realized by a HWP (H8a). Immediately following the HWP (H8b), a QWP (Q1) and a PBS are used to perform qubit measurements, and the two consecutive HWPs can be simplified to a single one (H8).

\subsection{Experimental realization of an arbitrary Werner state}
\label{App:WernerDecomposition-Any}

Although the decomposition that we have provided in Eq.~(2) in main text suffices for our experimental demonstration, it is not valid for $v>\frac{d+1}{2d}$, which leads to a negative value of $q=1-\frac{2d}{d+1}v$. For completeness, we provide below an alternative two-qudit decomposition of $\rhoWvd$ valid for all $v\in[0,1]$. Explicitly, we note that

\begin{subequations}\label{eq:Wernerstate_q2}
\begin{gather}
	W^{(d)}(q)=\frac{1}{n^d_-} \sum_{i<j} \xi^{i,j}(p,q),\qquad\\
	\xi^{i,j}(p,q) = q \xi_{2}^{i,j}+\left( 1-q \right) \left[ p \xi_{0}^{i,j}+\left( 1-p \right) \xi_{1}^{i,j} \right]
\end{gather}
\end{subequations}
where $p=\frac{(d+1)(1-v)}{d+1-2v}$, $q=\frac{2v}{d+1}$, $\xi_0^{i,j}=\varrho_0^{i,j} = \proj{\Psi^- _{i,j}}$, 
$\xi_1^{i,j}=\proj{\Psi^+ _{i,j}}$, $\ket{\Psi^\pm_{i,j}} = \frac{1}{\sqrt{2}}(\ket{i,j}\pm\ket{j,i})$,
$\xi_2^{i,j}=\varrho_1^{i,j}=\frac{\proj{ii}+\proj{jj}}{2}$.
It is easy to verify that $p,q \in [0,1]$ for $v\in[0,1]$, so all the Werner states can be prepared as a uniform mixture of the two-qubit states $\xi^{i,j}(p,q)$ acting locally on the qubit subspace spanned by $\ket{i}$ and $\ket{j}$. Accordingly, we can use the following optical setup to prepare the Werner state $\rhoWvd$.

\begin{figure}[htb]
\includegraphics[width=1\columnwidth]{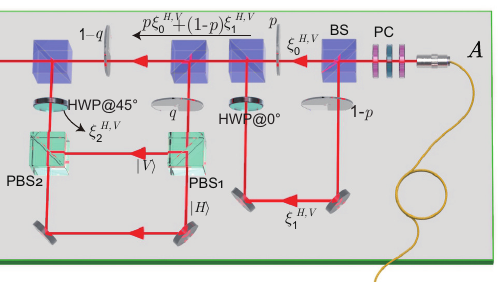}
\caption{Experimental setup at A for realizing $\xi^{i,j}$ in~\cref{eq:Wernerstate_q2}, which requires slight modifications to be made in the part of \cref{Fig:Setup_WernerPre1}.
}
\label{Fig:Setup_WernerPre1_02}
\end{figure}

\subsection{ScLFs beyond qubit projections} \label{App:ScLFS_Beyond_Qubit}

Although we have focused on the projection onto a qubit subspace in our experimental demonstration, it is worth noting that other ScLFs are possible with the path encoding of quantum information that we have employed. For example, suppose we have encoded our two-qudit quantum state $\rho$ using locally the $d$ path states $\{\ket{l_k}\}_{k=0}^{d-1}$ and we want to perform a local projection onto the subspace $\cH_\D=\text{span}\{\ket{l_l}\}_{k\in \D}$ where $\D\subset \bm{D}:=\{0,1,\cdots,d-1\}$, then it suffices to block the paths corresponding to $\bm{D}\setminus\D$. Then, if there are $d'$ elements in $\D$, such a filtering operation will transform locally the state space from dimension $d$ to dimension $d'$. In these notations, the specific ScLFs described in the main text simply correspond to $d=3$ and $\D=\{1,2\}$ with $d'=2$.

More generally, a local filtering operation transforming a $d$-dimensional state to a $d'$-dimensional state is described by a rectangular matrix $M$ of size $d'\times d$ such that $\| M\|_\infty = 1$ (i.e., having its maximum singular value equal to 1). Any such matrix $M$ admits the singular value decomposition
\begin{equation}\label{Eq:SVD}
	M = U\Sigma V^\dag
\end{equation}
where $V$ is a $d\times d$ unitary matrix, $\Sigma=\text{diag}(s_0,s_1,\cdots)$ is a $d'\times d$ matrix consisting of the singular values $s_i$ of $M$ on its diagonal, and $U$ is a $d'\times d'$ unitary matrix. To implement the local filtering operation corresponding to $M$, it then suffices to perform the following steps in sequence:
\begin{enumerate}
\item Apply the unitary transformation $V^\dag$ on the unfiltered space spanned by $\{\ket{l_k}\}_{k=0}^{d-1}$.
\item Apply the projection onto $\cH_\D$ where $\D$ consists of the collection of indices in $\bm{D}$ such that $s_i \neq 0 $ for all $i\in\D$. More precisely, in addition to blocking the paths corresponding to $\bm{D}\setminus\D$, signals propagating through path $i$ are also attenuated by the factor $s_i$.
\item Apply the unitary transformation $U$ on the filtered space spanned by $\{\ket{l_k}\}_{k\in\D}$.
\end{enumerate}

%\red{The experimental implementation of high-dimensional distillation, especially for large $d$, is still challenging with bulk optics. Integrated optics is a promising and scalable approach to generate $d$-dimensional entangled photons encoded in path~\cite{Wang2018Science,Bao2023NP}. We emphasize the distillation can be implemented on a single chip. Specifically, arbitrary unitary operations $U$ and $V$ can be implemented with $d^\prime$-channel and $d$-channel Mach-Zehnder
%interferometer~(MZI) network~\cite{Reck1994PRL,Clements2016Optica}. The blocking of path $i$, as well as the attenuation $s_i$ can be implemented by a programmable MZI, where the output can be thermally or electrically tuned. 

Note that the experimental preparation of good quality high-dimensional states and their distillation are still challenging with bulk optics, especially for large $d$. However, it is worth noting that integrated optics offers a promising and scalable approach to generate $d$-dimensional entangled photons encoded in the path degree of freedom~\cite{Wang2018Science,Bao2023NP}. In fact, the general single-copy local filtering operations discussed above can be implemented on a single chip in this physical implementation. More precisely, arbitrary unitary operations $U$ and $V$ of \cref{Eq:SVD} can be implemented with $d^\prime$-channel and $d$-channel Mach-Zehnder interferometer~(MZI) network~\cite{Reck1994PRL,Clements2016Optica}. The blocking of path $i$, as well as the attenuation $s_i$ can also be implemented by a programmable MZI, where the output can be thermally or electrically tuned.

\section{Miscellaneous Details and Theoretical Results}

\subsection{Fully-entangled fraction}\label{App:FEF}

In the main text, we claim the following Proposition
\begin{proposition}
If a {\em two-qubit} state $\rho$ satisfies $F_2>\frac{1}{2}$, it must also satisfy $F_d>\frac{1}{d}$ for all integer $d>2$.
\end{proposition}
In the following, we give a proof of this Proposition.
\begin{proof}
Let $\rho$ be a two-qubit state acting on the local qubit space $\cH_2=$ span$\{\ket{0},\ket{1}\}$ and $\ket{\Psi'_2}$ be the maximally entangled state (MES) saturating the fully-entangled fraction of $\rho$, i.e., 
\begin{equation}\label{Eq:F2}
	F_2(\rho) = \bra{\Psi'_2}\rho\ket{\Psi'_2} \quad\text{with}\quad\ket{\Psi'_2} =\Id_2\otimes U_2\ket{\Phi^+_2}
\end{equation} 
and $\ket{\Phi^+_d}:=\frac{1}{\sqrt{d}}\sum_{i=0}^{d-1}\ket{i}\ket{i}$. 

Now, define the $d\times d$ unitary operator $U_d:= U_2\oplus \Id_{d-2}$, where the identity operator $\Id_{d-2}$ acts on span$\{\ket{2},\ket{3},\cdots, \ket{d-1}\}$. Let $\ket{\Psi'_d}:=\Id_d\otimes U_d\ket{\Phi^+_d}$, then by embedding $\rho$ into the two-qudit space as $\tilde{\rho} = \rho\oplus {\bf 0}_{d^2-4}$, with the null operator ${\bf 0}_{d^2-4}$ acting on the subspace of $\mathbb{C}^d\otimes\mathbb{C}^d$ {\em orthogonal} to $\cH_2\otimes\cH_2$, we get
\begin{align*}
F_d(\rho) &= F_d(\tilde{\rho}):=\max_{\text{MES\ } \ket{\Psi_d}} \bra{\Psi_d}\tilde{\rho}\ket{\Psi_d} \\
& \ge \bra{\Psi'_d}\tilde{\rho}\ket{\Psi'_d} = \bra{\Phi^+_d} (\Id_d\otimes U_d^\dag)\tilde{\rho}(\Id_d\otimes U_d)\ket{\Phi^+_d}\\
&= \bra{\Phi^+_d} (\Id_d\otimes U_d^\dag)(\rho\oplus {\bf 0}_{d^2-4})\times\\
&\qquad\qquad\qquad\qquad\left[(\Id_2\oplus \Id_{d-2})\otimes (U_2\oplus \Id_{d-2})\ket{\Phi^+_d}\right]\\
&= \bra{\Phi^+_d} (\Id_d\otimes U_d^\dag)\left\{\left[\rho~(\Id_2\otimes U_2)\right]\oplus {\bf 0}_{d^2-4}\right\}\ket{\Phi^+_d}\\
&= \bra{\Phi^+_d} \left\{\left[(\Id_2\otimes U_2^\dag)\,\rho\,(\Id_2\otimes U_2)\right] \oplus {\bf 0}_{d^2-4}\right\} \ket{\Phi^+_d}\\
&= \frac{1}{d}\sum_{i,j=0}^{d-1}\bra{i}\bra{i}\!  \left\{\left[(\Id_2\otimes U_2^\dag)\,\rho\,(\Id_2\otimes U_2)\right]\! \oplus {\bf 0}_{d^2-4}\right\}\!\ket{j}\ket{j}\\
&= \frac{1}{d}\sum_{i,j=0}^{1}\bra{i}\bra{i}\!\left[(\Id_2\otimes U_2^\dag)\,\rho\,(\Id_2\otimes U_2)\right]\! \ket{j}\ket{j}\\
&= \frac{2}{d} \bra{\Phi^+_2}(\Id_2\otimes U_2^\dag)\,\rho\,(\Id_2\otimes U_2) \ket{\Phi^+_2}\\
&= \frac{2}{d} \bra{\Psi_2'}\rho\ket{\Psi_2'}>\frac{2}{d}\times \frac{1}{2} = \frac{1}{d}
\end{align*}
where we have used in the last line \cref{Eq:F2} and the premise $F_2(\rho)>\frac{1}{2}$.
\end{proof}

\subsection{Bell-nonlocality}\label{App:BellNonlocality}

\subsubsection{Symmetric (quasi-)extension}
\label{App:SymExt}

A powerful way to determine if a quantum state $\rhoAB$ could violate any Bell inequality is to check for the extent to which it admits a symmetric 
(quasi-)extension~\cite{DPS:PRL:2022,Terhal03}. In particular, if $\rhoAB$ has a $(s_a,1)$-symmetric (quasi-) extension, then it {\em cannot} violate any Bell inequality that has $s_a$ settings for Alice, and an {\em arbitrary} number of settings for Bob. A similar statement holds for a quantum state admitting a $(1,s_b)$-symmetric (quasi-) extension.

For completeness (and simplicity of presentation), we provide below a semidefinite program (SDP) that facilitates the search for the existence of a $(k,1)$-symmetric extension~\cite{DPS:PRL:2022} for any given $\rhoAB$ acting on $\mathbb{C}^{d_A}\otimes\mathbb{C}^{d_B}$. For the rest of this subsection, we shall refer to any $\rhoAB$ admitting a $(k,1)$-symmetric extension as being $k$-extendible. Our SDP formulation differs somewhat from that of~\cite{DPS:PRL:2022,Terhal03} but makes explicit use of the two facts: (1) the set of $k$-extendible density matrices $\rhoAB$ is {\em convex}, and (2) the maximally mixed state $\frac{\Id_D}{D}$, where $D:=d_Ad_B$ is $k$-extendible for all $k>1$. Hence, if mixing $\rhoAB$ with a negatively-weighted $\frac{\Id_D}{D}$ is $k$-extendible, $\rhoAB$ itself must already be $k$-extendible:

\begin{align}\label{Eq:SDP:SymmExt}
& \ \ \min_{t\ge0,\, \Tilde{\rho}\succeq 0}  \qquad\qquad\qquad\qquad t \\
    &\text{such that}\quad \tr_{\mathcal{H}^{\Bar{\kappa}_i}}{\Tilde{\rho}}=\rhoAB+(t-1) \frac{\Id_D}{D},~\forall\,\, \kappa_i\in \{1,\dots,k\},\nonumber
\end{align}
where $\Bar{\kappa}_i=\{1,\dots,k\}\backslash \kappa_i$, i.e., the partial trace is over all but the $i$-th copy of $\mathbb{C}^{d_A}$ and   $\succeq 0$ means matrix positivity. If the optimum value of this optimization is $t=t^\star\le 1$, then $\Tilde{\rho}+(1-t^\star)\frac{\Id_{d_A^kd_B}}{d_A^kd_B}$ is a legitimate quantum state that serves as the $(k,1)$-symmetric extension of $\rhoAB$.\footnote{Generalization to the search for a $(s_a,s_b)$-symmetric extension of $\rhoAB$ for any positive integers $s_a,s_b\ge 1$ should be evident from \cref{Eq:SDP:SymmExt}.}

Consider now $\rhoAB(w)$ decomposable as a convex mixture of an entangled state $\sigma$ and the maximally mixed state $\frac{\Id_D}{D}$:
\begin{equation}\label{Eq:rhoAB:Decomposition}
	\rhoAB(w)=(1-w)\sigma + w\frac{\Id_D}{D}.
\end{equation}
 Let $t^\star_0$ be the optimum value of $t$ obtained by solving \cref{Eq:SDP:SymmExt} for $\sigma:=\rhoAB(w=0)$ 
 and $w=w_t$ be the critical weight above which $\rhoAB(w)$ becomes $k$-extendible. 
By definition and the constraints of \cref{Eq:SDP:SymmExt}, we see that upon normalization, 
\begin{equation}\label{Eq:Critical}
	\frac{1}{\tsz}\left[\sigma + (\tsz-1)\frac{\Id_D}{D}\right]=\rhoAB\left(\frac{\tsz-1}{\tsz}\right)
\end{equation}
must be $k$-extendible. In particular, all $\rhoAB(w)$ with $w<\frac{\tsz-1}{\tsz}$ is {\em not} $k$-extendible, or it would contradict the assumption that $\tsz$ is the optimum value obtained by solving \cref{Eq:SDP:SymmExt} for $\rhoAB=\sigma$. Hence, $w_t = \frac{\tsz-1}{\tsz}$.

\begin{table*}%[h!]
    \begin{tabular}{|c||c|c|c|c|}
    \hline
    $(1,2)$& \multicolumn{2}{|c|}{$\rhoExp$}& \multicolumn{2}{|c|}{$\rhoWv{3}$}\\
    \hline 
    $v$ & SE & SQE & SE & SQE \\
    \hline 
   $0$ & 1.0131   & 1.0086 & 1 & 1  \\
   \hline 
   $0.05$ & 1.0017   & 0.9876 & 0.9250 & 0.9250  \\
   \hline 
   $0.1$ & 1.0003   & 0.9886 & 0.8500 & 0.8500  \\
   \hline 
   $0.15$ & 0.9670   & 0.9369  & 0.7750 & 0.7750 \\
   \hline 
   $0.2$ & 0.9145   & 0.8469 & 0.7000 & 0.7000  \\
   \hline 
   $0.25$ & 0.7758   & 0.7022 & 0.6250 & 0.6250  \\
   \hline 
   $0.3$ & 0.7392   & 0.7004  & 0.5500 & 0.5500 \\
   \hline 
   $0.35$ & 0.5926   & 0.5505  & 0.4750 & 0.4750 \\
   \hline 
   $0.4$ & 0.6649   & 0.6161 & 0.4000 & 0.4000  \\
   \hline 
   $0.45$ & 0.4460   & 0.4062  & 0.3250 & 0.3250 \\
   \hline 
    \end{tabular}
    \quad
    \begin{tabular}{|c||c|c|c|c|}
    \hline
    $(1,3)$ & \multicolumn{2}{|c|}{$\rhoExp$}& \multicolumn{2}{|c|}{$\rhoWv{3}$}\\
    \hline 
    $v$ & SE & SQE & SE & SQE \\
    \hline 
   $0$ & 1.2725   & 1.2725  & 1.3333 & 1.3333 \\
   \hline 
   $0.05$ & 1.1813   & 1.1813 & 1.2333 & 1.2333  \\
   \hline 
   $0.1$ & 1.1630   & 1.1629  & 1.1333 & 1.1333 \\
   \hline 
   $0.15$ & 1.0939   & 1.0938  & 1.0333 & 1.0333 \\
   \hline 
   $0.2$ & 0.9912   & 0.9850  & 0.9333 & 0.9333 \\
   \hline 
   $0.25$ & 0.8387   & 0.8366  & 0.8333 & 0.8333 \\
   \hline 
   $0.3$ & 0.8083   & 0.8060  & 0.7333 & 0.7333 \\
   \hline 
   $0.35$ & 0.6359   & 0.6326  & 0.6333 & 0.6333 \\
   \hline 
   $0.4$ & 0.6668   & 0.6524  & 0.5333 & 0.5333 \\
   \hline 
   $0.45$ & 0.4593   & 0.4510  & 0.4333 & 0.4333 \\
   \hline 
    \end{tabular}
    \quad
    \begin{tabular}{|c||c|c|c|c|}
    \hline
    $(1,4)$ & \multicolumn{2}{|c|}{$\rhoExp$}& \multicolumn{2}{|c|}{$\rhoWv{3}$}\\
    \hline 
    $v$ & SE & SQE & SE & SQE \\
    \hline 
   $0$ &  1.5103  & 1.5098 & 1.6000 & 1.6000   \\
   \hline 
   $0.05$ & 1.3688   & 1.3685 & 1.4800 & 1.4800  \\
   \hline 
   $0.1$ &  1.3368  & 1.3363 & 1.3600 & 1.3600  \\
   \hline 
   $0.15$ &  1.2661  & 1.2659 & 1.2400 & 1.2400 \\
   \hline 
   $0.2$ &  1.1248  & 1.1222 & 1.1200 & 1.1200 \\
   \hline 
   $0.25$ & 0.9614   & 0.9610 & 1.0000 & 1.0000  \\
   \hline 
   $0.3$ & 0.9243   & 0.9225 & 0.8800 & 0.8800 \\
   \hline 
   $0.35$ & 0.7237   & 0.7221 & 0.7600 & 0.7600 \\
   \hline 
   $0.4$ & 0.7118   & 0.7111 & 0.6400 & 0.6400 \\
   \hline 
   $0.45$ & 0.5017   & 0.5004  & 0.5200 & 0.5200  \\
   \hline 
    \end{tabular}
    \\ \vspace{0.2cm}
    \begin{tabular}{|c||c|c|c|c|}
    \hline 
    $(2,1)$ & \multicolumn{2}{|c|}{$\rhoExp$}& \multicolumn{2}{|c|}{$\rhoWv{3}$}\\
    \hline 
    $v$ & SE & SQE & SE & SQE\\
    \hline
   $0$ & 1.0196   & 1.0159  & 1 & 1  \\
   \hline 
   $0.05$ &1.0000   & 0.9876 & 0.9250 & 0.9250  \\
   \hline 
   $0.1$ &1.0000   & 0.9886  & 0.8500 & 0.8500 \\
   \hline 
   $0.15$ &0.9670   & 0.9369 & 0.7750 & 0.7750  \\
   \hline 
   $0.2$ &0.9145   & 0.8469  & 0.7000 & 0.7000 \\
   \hline 
   $0.25$ &0.7758   & 0.7023 & 0.6250 & 0.6250  \\
   \hline 
   $0.3$ &0.7392   & 0.7004 & 0.5500 & 0.5500  \\
   \hline 
   $0.35$ &0.5926   & 0.5505 & 0.4750 & 0.4750  \\
   \hline 
   $0.4$ &0.6649   & 0.6161  & 0.4000 &  0.4000\\
   \hline 
   $0.45$ &0.4460   & 0.4062 & 0.3250 & 0.3250  \\
   \hline 
    \end{tabular}
    \quad
    \begin{tabular}{|c||c|c|c|c|}
    \hline
    $(3,1)$ & \multicolumn{2}{|c|}{$\rhoExp$}& \multicolumn{2}{|c|}{$\rhoWv{3}$}\\
    \hline 
    $v$ & SE & SQE & SE & SQE \\
    \hline 
   $0$ & 1.2753   & 1.2753 & 1.3333 & 1.3333   \\
   \hline 
   $0.05$ & 1.1702   & 1.1700 & 1.2333 & 1.2333  \\
   \hline 
   $0.1$ & 1.1478   & 1.1472 & 1.1333 & 1.1333 \\
   \hline 
   $0.15$ & 1.0839   & 1.0830 & 1.0333 & 1.0333 \\
   \hline 
   $0.2$ & 0.9842   & 0.9787 & 0.9333 & 0.9333 \\
   \hline 
   $0.25$ & 0.8423   & 0.8401  & 0.8333 & 0.8333 \\
   \hline 
   $0.3$ & 0.8083   & 0.8063 & 0.7333 & 0.7333 \\
   \hline 
   $0.35$ & 0.6359   & 0.6319  & 0.6333 & 0.6333 \\
   \hline 
   $0.4$ & 0.6680   & 0.6553 & 0.5333 & 0.5333  \\
   \hline 
   $0.45$ & 0.4600   & 0.4508  & 0.4333 & 0.4333 \\
   \hline 
    \end{tabular}
    \quad
    \begin{tabular}{|c||c|c|c|c|}
    \hline
    $(4,1)$ & \multicolumn{2}{|c|}{$\rhoExp$}& \multicolumn{2}{|c|}{$\rhoWv{3}$}\\
    \hline 
    $v$ & SE & SQE & SE & SQE \\
    \hline 
   $0$  &  1.5097  & 1.5094 & 1.6000 & 1.6000  \\
   \hline 
   $0.05$ & 1.3660   & 1.3655 & 1.4800 & 1.4800  \\
   \hline 
   $0.1$ & 1.3321   & 1.3311 & 1.3600 & 1.3600 \\
   \hline 
   $0.15$ & 1.2633   & 1.2631 & 1.2400 & 1.2400 \\
   \hline 
   $0.2$ &  1.1237  & 1.1197 & 1.1200 & 1.1200 \\
   \hline 
   $0.25$ & 0.9628   & 0.9621 & 1.0000 & 1.0000 \\
   \hline 
   $0.3$ &  0.9223  & 0.9209 & 0.8800 & 0.8800 \\
   \hline 
   $0.35$ & 0.7217   & 0.7205 & 0.7600 & 0.7600 \\
   \hline 
   $0.4$ & 0.7139   & 0.7134 & 0.6400 & 0.6400  \\
   \hline 
   $0.45$ & 0.5012   & 0.5001 & 0.5200 & 0.5200 \\
   \hline 
    \end{tabular}
\caption{Optimum results $t^\star$ obtained by solving~\cref{Eq:SDP:SymmExt} [\cref{Eq:SDP:QSymmExt}] for the existence of a $(k,1)$ or a $(1,k)$-symmetric (quasi-)extension, abbreviated as SE (SQE), for the ideal two-qutrit Werner state $\rhoWv{3}$ and its experimental counterpart $\rhoExp$. The state has a symmetric (quasi-)extension \Iff $t^\star\le1$. Due to limitations in the computational resource, in our computation of the symmetric quasi-extension for $k=4$ copies, only a subset of bipartitions is considered in the sum in~\cref{Eq:SDP:QSymmExt}. See text for details. Since a Werner state is invariant under swapping Alice and Bob, i.e., $\rhoWvd = V\rhoWvd V^\dag$, we see that $\rhoWvd$ has a $(k,1)$ -symmetric (quasi-)extension \Iff it also has a $(1,k)$-symmetric (quasi-)extension. The experimentally prepared states $\rhoExp$, however, are not exactly invariant under this permutation, resulting in slightly different values of $t^\star$ when solving for the symmetric (quasi-)extension with respect to the two parties.} 
\label{TABLE:SQE}
\end{table*}

For Werner states $W^{(d)}(w)$, using the identity $\Pi^d_+=\Id-\Pi^d_-$, we can also rewrite it from Eq.~(1) in main text  as 
\begin{equation}
	W^{(d)}(v) = \left(1-\frac{D v}{n^d_+}\right)\frac{\Pi^d_-}{n^d_-} + \frac{D v}{n^d_+}\frac{\Id}{D},
\end{equation}
which is in the form of \cref{Eq:rhoAB:Decomposition} if
$\sigma = \frac{\Pi^d_-}{n^d_-}$, $w=\frac{D v}{n^d_+}$, then
\begin{equation}\label{Eq:vt}
	 v_t=\frac{n^d_+}{D}w_t=\frac{n^d_+}{D}\frac{\tsz-1}{\tsz},
\end{equation}
i.e., $\rhoWvd$ for all $v>v_t$ is $k$-extendible.

To rule out the possibility of $\rhoAB$ violating any Bell inequality with a restricted number of settings, one can relax the SDP of \cref{Eq:SDP:SymmExt} by considering an extension that is an entanglement witness, and hence {\em not} positive semidefinite. In particular, if we consider only a decomposable entanglement witness, then the following SDP facilitates the search for a symmetric $(k,1)$ quasi-extension of $\rhoAB$:
\begin{align}\label{Eq:SDP:QSymmExt}
& \ \ \min_{t\ge0}  \qquad\qquad\qquad\qquad t \nonumber\\
    &\text{such that}\quad \Tilde{H}_\rho = P+\sum_{p\in \mathcal{P}} Q^{\text{T}_p}_p,\nonumber\\
    & \qquad\qquad\ \  P\succeq 0,\quad Q_p \succeq 0 ~\forall\,\, p \in \mathcal{P}\\
    &\qquad\qquad\ \  \tr_{\mathcal{H}^{\Bar{\kappa}_i}}{\Tilde{H}_p}=\rhoAB+(t-1) \frac{\Id_D}{D},~\forall\,\, \kappa_i\in \{1,\dots,k\},\nonumber
\end{align}
where $\mathcal{P}$ is the set of all possible bipartitions. As with \cref{Eq:SDP:SymmExt}, if the optimum value of \cref{Eq:SDP:QSymmExt}  $t=t^\star\le 1$, then $\Tilde{H}_p+(1-t^\star)\frac{\Id_{d_A^kd_B}}{d_A^kd_B}$ serves as the $(k,1)$-symmetric quasiextension of $\rhoAB$. 

In \cref{TABLE:SQE}, we list our numerical results obtained by solving ~\cref{Eq:SDP:SymmExt} and~\cref{Eq:SDP:QSymmExt} 
for the Werner states $\rhoWvd$. Note that due to limitations of computational resources, for determining the $(4,1)$-symmetric quasiextension, we only consider in $\mathcal{P}$ the following subset of bipartitions: $\{A_1|A_2A_3A_4B\}, \{B|A_1A_2A_3A_4\},\{A_1A_2|A_3A_4B\}$, and $\{A_1B|A_2A_3A_4\}$, where $A_k$ denotes the $k$-copy of Alice's subsystem. A similar simplification for the computation related to the $(1,4)$-symmetric quasiextension is adopted. Our findings suggest that in the case of Werner states $\rhoWvd$, the relaxation from~\cref{Eq:SDP:SymmExt} to~\cref{Eq:SDP:QSymmExt} (with the simplification considered above) does {\em not} lead to the construction of a LHVM for a wider interval of entangled Werner states. 

\begin{table*}%[h!]
    \centering
    \begin{tabular}{|c||cc||cc||cc||cc||cc||cc||cc||cc||cc||cc||cc||cc|}
    \hline
    & \multicolumn{2}{c||}{(1,2)}& \multicolumn{2}{c||}{(1,3)}& \multicolumn{2}{c||}{(1,4)}& \multicolumn{2}{c||}{(1,5)}& \multicolumn{2}{c||}{(1,6)}& \multicolumn{2}{c||}{(1,7)}& \multicolumn{2}{c||}{(1,8)}& \multicolumn{2}{c||}{(1,9)}& \multicolumn{2}{c||}{(1,10)}& \multicolumn{2}{c||}{(1,11)}& \multicolumn{2}{c||}{(1,12)}& \multicolumn{2}{c|}{(1,13)}\\
    \hhline{|=::=:=::=:=::=:=::=:=::=:=::=:=::=:=::=:=::=:=::=:=::=:=::=:=|}
       $d$ & SE & SE-B & SE & SE-B& SE & SE-B& SE & SE-B& SE & SE-B& SE & SE-B& SE & SE-B& SE & SE-B& SE & SE-B& SE & SE-B& SE & SE-B& SE & SE-B\\
    \hline
    $2$ & $\frac{1}{4}$ & $\frac{1}{4}$ & $\frac{1}{3}$ & $\frac{1}{3}$ & $\frac{3}{8}$ & $\frac{3}{8}$ & $\frac{2}{5}$ & $\frac{2}{5}$ & $\frac{5}{12}$ & $\frac{5}{12}$ & $\frac{3}{7}$ & $\frac{3}{7}$ &   & $\frac{7}{16}$ &  & $\frac{4}{9}$ &  & $\frac{9}{20}$ &  & $\frac{5}{11}$ & & $\frac{11}{24}$& & $\frac{6}{13}$\\
    
    $3$ & $0$ & $\frac{1}{4}$ & $\frac{1}{6}$ & $\frac{1}{3}$ & $\frac{1}{4}$ & $\frac{3}{8}$ & & $\frac{2}{5}$ & & $\frac{5}{12}$ & & $\frac{3}{7}$ & & $\frac{7}{16}$ &&&&&&&&&&\\
    $4$ & $0$ & $\frac{1}{4}$ & $0$ & $\frac{1}{3}$ &  & $\frac{3}{8}$ &   & $\frac{2}{5}$ & & $\frac{5}{12}$ & & & & & & & & & & & & & & \\
    $5$ & $0$ & $\frac{1}{4}$ &  & $\frac{1}{3}$ & & $\frac{3}{8}$ &  &  & & & & & & & & & & & & & & & & \\
    $6$ & $0$ & $\frac{1}{4}$ & & $\frac{1}{3}$& & & & & & & & & & & & & & & & & & & & \\
    $7$ & $0$ & $\frac{1}{4}$ & &  & & & & & & & & & & & & & & & & & & & & \\
    \hline
    \end{tabular}
    \caption{Threshold value of $v$, denoted by $v_t$, above which the $d$-dimensional Werner states $\rhoWvd$ are found to admit a $(1,k)$ symmetric extension (SE) or a $(1,k)$ {\em bosonic} symmetric extension (SE-B)~\cite{Doherty:2014aa}. The latter is a special kind of SE where the extension $\Tilde{\rho}$ is further required to reside in the symmetric subspace of the $k$ copies of Alice. Being more constraining, the existence of a SE-B represents a tighter separability criterion compared with the existence of a SE. That is, for any given $d$ and $k$, the value of $v_t$ for SE-B is larger than or equal to that for SE.  The values of $v_t$ presented in this Table may be related to the values of $t^\star$ presented in \cref{TABLE:SQE} for $W^{(d)}(v=0)$ via \cref{Eq:vt}. 
    Numerically, a $(1,k)$ extension is found until $v\simeq v_t-10^{-4}$.  
    Blank entries correspond to those situations where we could not solve the SDP of~\cref{Eq:SDP:SymmExt} using the computation resources available. The value of $v_t=v_\text{Sym}:=\frac{1}{2}(1-\frac{d-1}{k})$ corresponding to the $(1,k)$ SE of $\rhoWvd$ has been determined independently and analytically in~\cite{JV23} (see also~\cite{Terhal03}). Interestingly,  our results determined using the QETLAB toolbox~\cite{qetlab} suggests that $\rhoWvd$ has a $(1,k)$ SE-B whenever $v\ge \frac{1}{2}\left(1-\frac{1}{k}\right)=\left.v_\text{Sym}\right|_{d=2} \ge v_\text{Sym}$, i.e., a $(1,k)$ SE-B for $\rhoWvd$ exists for all $d>2$ \Iff a SE for $\rhoWv{d=2}$ exists.}
    \label{tab:my_label}
\end{table*}

In \cref{tab:my_label}, we also show, cf. \cref{Eq:vt}, the threshold value of the symmetric weight above which the Werner states $\rhoWvd$ admit a $(1,k)$ (bosonic~\cite{Doherty:2014aa}) symmetric extension.

From \cref{TABLE:SQE}, we see that the tomographically-reconstructed experimentally states $\rhoExp$ more or less exhibit the same level of $k$-(quasi)extendability for $k\le 4$. For larger values of $k$, we remark that by using an adaptation~\cite{Note2-Yujie} of the method from~\cite{Villegas-Aguilar:2024aa}, we could verify that for $v=0.45$, a convex decomposition of the following kind can be obtained for the experimentally produced state:
\begin{equation}\label{Eq:Decompose}
\begin{split}
	\rhoExp &= p\,\rhoWv{3} + \sum_i q_i\,\id\otimes U_i W^{(3)}(v)\id\otimes U_i^\dag\\	
	&+\sum_{\ell, j,m} r_{\ell jm}\,V_\ell\otimes U_\ell \proj{j}\otimes\proj{m}\,V_\ell^\dag\otimes U_\ell^\dag,
\end{split}
\end{equation}
where $\{\ket{j}\}$ and $\{\ket{m}\}$ are computational basis states for the first and second Hilbert spaces, $V_\ell$ and $U_\ell$ are some unitaries acting on the two Hilbert space, $p,q_i,r_{\ell jm}\ge0$ for all $i,\ell,j,m$, and $p+\sum_i p_i+\sum_{\ell jk} r_{\ell jm} =1$. Notice that each term in the first sum on the RHS of~\cref{Eq:Decompose} has exactly the same (non)local property as $\rhoWv{3}$ while the second sum clearly gives a separable state, which is $k$-(quasi)extendable for {\em all} integers $k\ge 1$. Using the convexity of the set of $k$-(quasi)extendable states (for any fixed but arbitrary $k$), we thus see that if $\rhoWv{3}$ is  $k$-(quasi)extendable for some $k$, so is the corresponding $\rhoExp$. 

Note further that unlike the case of two-qubit states discussed in~\cite{Villegas-Aguilar:2024aa}, the partial-transposition criterion~\cite{HHHH09} is only necessary but {\em not} sufficient for separability for the higher-dimensional systems we are considering. Consequently, in searching for a decomposition like \cref{Eq:Decompose} for a given $v$, we settle for a restricted set of separable states given by the second sum of \cref{Eq:Decompose} (facilitated by a randomly sampled set of unitaries), rather than optimizing over all separable two-qutrit states.

\subsubsection{Finding Bell-nonlocal correlations for the experimentally produced two-qutrit states}
\label{App:FindBellViolation}

For any given quantum correlation $\vec{P}_\Q$ attained by performing the local measurements described by the positive-operator-valued measures (POVMs) $\{M^A_{a|x}\}_{a,x}$ and $\{M^B_{b|y}\}_{b,y}$ on a shared density matrix $\rho_\text{AB}$:
\begin{equation}
	P_\Q(a,b|x,y) = \tr(\rho_\text{AB}\,M^A_{a|x}\otimes M^{B}_{b|y}),
\end{equation}
we can determine if it is Bell-local by determining its nonlocal content~\cite{EPR92}:
\begin{subequations}
\begin{gather}
\min_{v\in[0,1],\vec{P}\in\mathcal{L}, \vec{P}_\text{NS}}\qquad\quad v\\
\text{such that \ \ } (1-v)\vec{P} + v\vec{P}_\text{NS}=\vec{P}_\Q
\end{gather}
\end{subequations}
where $\vec{P}_\text{NS}$ is any non-signaling correlation and $\mathcal{L}$ is the set of Bell-local correlations. If the result of this optimization is larger than $0$,  we can conclude that $\vec{P}_\Q\not\in\mathcal{L}$ and hence $\vec{P}$ is Bell-nonlocal.

Since all $\rhoExp$ with $v=0.01,\ldots, 0.45$ admit a 2-symmetric quasiextension, it is necessary to consider a Bell scenario with at least $n_s=3$ measurement settings per site to identify its potential Bell-nonlocality. Moreover, given that these are two-qutrit states, it seems natural to consider Bell scenarios involving $n_o=3$ outcomes. To this end, our numerical maximizations using $\rhoExp$ and the LB (see-saw) algorithm explained in~\cite{LiangPRA07} have not led to any violation of the $21,169$ classes of facet-defining Bell inequalities from~\cite{CopePRA19}. In addition, we perform around $30,000$ {\em double} optimizations over the local measurements $\{M^A_{a|x}\}_{a,x}$ and $\{M^B_{b|y}\}_{b,y}$ to maximize the nonlocal content associated with these $\rhoExp$ (including the one for $v=0$) for the $(n_s,n_o)=(3,3)$ Bell scenario and several thousand optimizations for the $(n_s,n_o)=(4,3)$. In all these cases, we have {\em not} found a single instance of $\vec{P}_\Q$ where the corresponding nonlocal content $v$ is unmistakably larger than zero.

\subsection{Steering robustness}
\label{App:SR}

When Alice performs local measurements $\{M^A_{a|x}\}_{a,x}$ on the shared state $\rhoAB$, quantum theory dictates that, up to normalization, the conditional state prepared at Bob's side is:
\begin{equation}
    \rho_{a|x} = \tr_A\left[{(M^A_{a|x}\otimes \Id)\rho_{\text{AB}}}\right].
\end{equation} 
In the studies of quantum steering~\cite{UolaRMP20}, the collection of these subnormalized conditional density matrices $\{\rho_{a|x}\}_{a,x}$ is called an assemblage. Accordingly, the extent to which Alice's measurements can steer this assemblage can be quantified using the so-called steering robustness SR~\cite{PianiPRL15}, defined as the minimum value of $t$ such that $\{\frac{\rho_{a|x}+t\tau_{a|x}}{1+t}\}_{a,x}$ is unsteerable (admitting a local-hidden-state model) where $\{\tau_{a|x}\}_{a,x}$ is a valid assemblage.

For a given assemblage $\{\rho_{a|x}\}_{a,x}$, the corresponding SR can also be computed by solving the following SDP~\cite{2017_cavalcanti_Rep.Prog.Phys._Quantum}:
\begin{equation}
    \begin{split}
        \min_{\{\sigma_\lambda\}} & \quad \left( \sum_\lambda \tr\, \sigma_\lambda\right)-1\\
        \text{s.t.} & \quad \sum_{\lambda}D(a|x,\lambda)\sigma_\lambda \succeq \rho_{a|x}~\forall a,x\\
        & \quad \sigma_\lambda\succeq 0 ~\forall\,\, \lambda,
    \end{split}
\end{equation}
where $D(a|x,\lambda)=\delta_{a,\lambda_x}$ are deterministic response functions and $\lambda = (\lambda_1,\dots,\lambda_{n_s})$.

To quantify the steerability of a quantum state $\rho_{\text{AB}}$ from A to B, one can optimize $\SR$ over all  possible measurement strategies of Alice, i.e.,
\begin{align}\label{Eq:SR:AtoB}
        \SR_{A\to B}\left(\rho_{\text{AB}}\right):= &\max_{\{M^A_{a|x}\}_{a,x}}\quad \SR\left(\{\rho_{a|x}\}_{a,x}\right)\\
        &\text{s.t.}\quad \rho_{a|x} =  \tr_A\left[{(M^A_{a|x}\otimes \Id)\rho_{\text{AB}}}\right]~\forall\,\, a,x.\nonumber
\end{align}
Evidently, a similar definition can be given for $\SR_{B\to A}\left(\rho_{\text{AB}}\right)$. Then, following~\cite{SLChen16}, we can define 
\begin{equation}\label{Eq:SR}
	\SR\left(\rho_{\text{AB}}\right):=\max\{ \SR_{A\to B}\left(\rho_{\text{AB}}\right), \SR_{B\to A}\left(\rho_{\text{AB}}\right)\}.
\end{equation}
\begin{figure}[h!t]
	\includegraphics[width=\columnwidth]{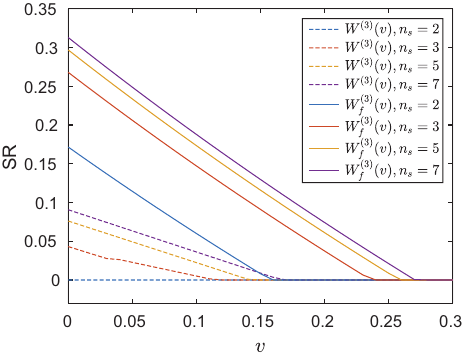}
	\caption{Best lower bounds on the steering robustness $\SR$ found for $\rhoWv{3}$ (dashed lines) and $\rhoWf{3}$ (solid lines) for various number $n_s$ of measurement settings and $v=0, 0.01, \cdots, 0.3$.}
	\label{fig:SR_W}
\end{figure}
For $\rhoWvd$ and $\rhoWfd$, both invariant with respect to swapping $A$ and $B$, we have  $\SR\left(\rho_{\text{AB}}\right)= \SR_{A\to B}\left(\rho_{\text{AB}}\right)$. In principle, there is no limitation to the number of measurement settings $n_s$ and the number of outcomes $n_o$ in~\cref{Eq:SR:AtoB}. In \cref{fig:SR_W}, we provide the best lower bounds on $\SR$ for $\rhoWv{3}$ and $\rhoWf{3}$ we have found by considering up to $n_s = 7$. For each of them, we set $n_o=d$, the local Hilbert space dimension, and obtain these results using the optimization code of~\cite{2017_cavalcanti_Rep.Prog.Phys._Quantum} by considering $1,000$ random initial local measurements $\{M^A_{a|x}\}_{a,x}$. It is worth noting from~\cref{fig:SR_W} that with only $n_s=2$ measurement settings, it appears {\em impossible} to demonstrate the steerability of $\rhoWv{3}$.

\subsection{Dense-codability}
\label{App:DenseCodability}

The Holevo bound~\cite{M.A.Nielsen:Book:2000} says that the amount of classical information that can be transmitted using a $d$-dimensional quantum state is at most $\log_{2}(d)$ bits. However, with shared entanglement, it is possible to go beyond this bound through a dense coding protocol~\cite{CW92}. To this end, let Alice and Bob share a two-qudit state $\rhoAB$. Suppose Alice applies a local unitary $U_i$ with probability $p_i$ on her qudit and sends it to Bob. This means that she effectively prepares an ensemble $\{p_i, \rho_{\text{AB},i}\}$ for Bob. Upon receiving the qudit, Bob tries to decode the index $i$ by performing a joint measurement on $(U_i\otimes\Id_B)\rhoAB(U_i\otimes\Id_B)^\dag$.

By using an optimal choice of orthogonal unitaries, Ref.~\cite{Bruss2004densecoding} shows that the dense coding capacity $\chi$ (in bits) of a two-qudit state $\rhoAB$ with Alice sending her qudit to Bob is quantified by $\chi(\rhoAB) := \log_{2}d + S(\rho_\text{B}) - S(\rhoAB)$, where $S(\cdot)$ denotes the von Neumann entropy~\cite{M.A.Nielsen:Book:2000}. 
Compared with the Holevo bound $\log_2d$, one finds that a quantum state is useful for dense coding from A to B if %the quantity
\begin{equation}
    \delta(\rhoAB) := S(\rho_\text{B}) - S(\rhoAB) > 0.
\end{equation}

For the two-qudit Werner states $\rhoWvd$, we have
\begin{equation}\label{Eq:WernerDelta}
\begin{aligned}
    \delta[\rhoWvd] = &\log_2d + v\log_2 \frac{2v}{d(d+1)}  \\
    &+ (1-v)\log_2\frac{2(1-v)}{d(d-1)},
\end{aligned}
\end{equation}
which is easily verified to be negative for all $d\ge 3$ and $v \in [0,1]$. Thus, all $\rhoWvd$ for $d\ge 3$ 
are useless for dense coding.

\begin{figure}[h!t]
	\includegraphics[width=\columnwidth]{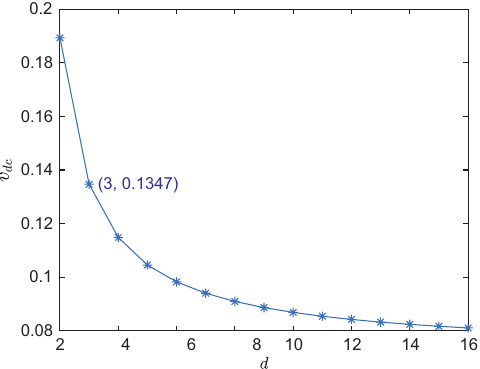}
	\caption{Threshold value of $v$ below which \cref{Eq:DeltaFiltered} becomes larger than zero, i.e., where $\rhoWvd$ may be locally filtered by qubit projection to become useful for dense coding.}
	\label{fig:DC_W}
\end{figure}

However, upon qubit filtering operation, $\rhoWvd$ becomes  $\rhoWfd$, which is local-unitarily equivalent to $W^{(2)}(v')$ with $v'=\frac{3(d-1)v}{N_{d,v}}$ where $N_{d,v}:= (d+1)(1-v) + 3v(d-1)$. Using this in~\cref{Eq:WernerDelta} with $d=2$, we obtain
\begin{align}\label{Eq:DeltaFiltered}
    \delta[\rhoWfd] = &~1 + \frac{(d+1)(1-v)}{N_{d,v}}\log_2\frac{(d+1)(1-v)}{N_{d,v}} \nonumber \\
    & + \frac{3v(d-1)}{N_{d,v}}\log_2\frac{v(d-1)}{N_{d,v}}. 
\end{align}
Numerically, we can solve the critical value of $v$, denoted by $v_{dc}$ below which $\delta[\rhoWfd]$ becomes larger than zero. The corresponding results for $d=2,3,\cdots, 16$ are shown in \cref{fig:DC_W}. In the asymptotic limit of $d\to\infty$, our computation suggests that $v_{dc}$ approaches $\sim 0.0722088$.

\section{Other experimental results and analysis}

Here, we provide other relevant experimental results omitted from the main text, including the reconstructed density matrices, their fidelity with respect to the reference state,  the steering robustness, and the dense codability of the filtered quantum states.

\subsection{Experimental states and fidelity}\label{App:Fidelity}

\begin{figure*}[h!tb]
	\includegraphics[scale=1]{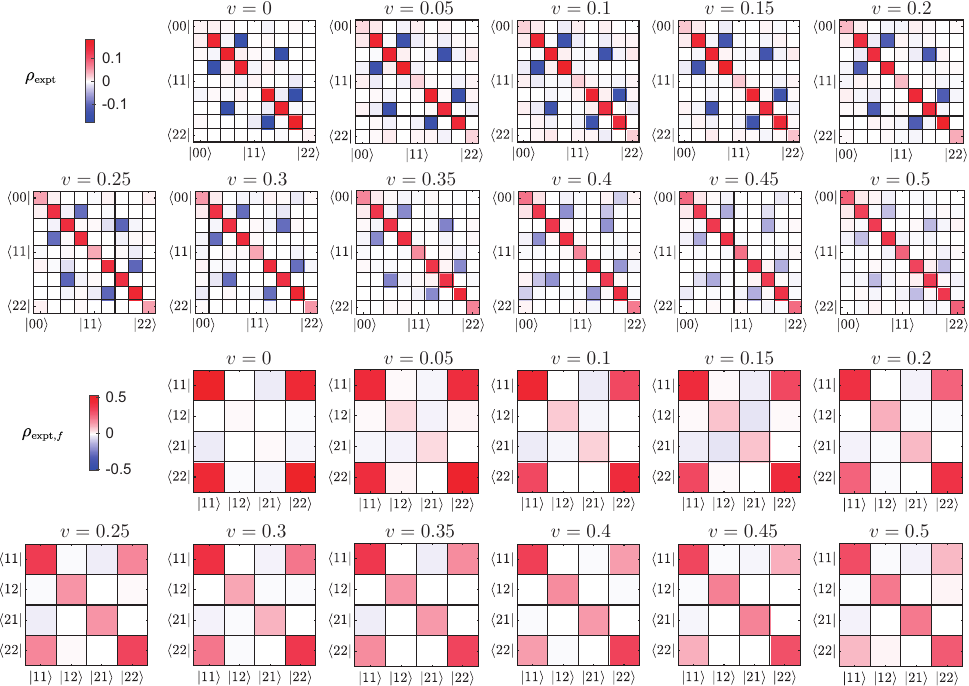}
	\caption{Experimentally (reconstructed) density matrix of the unfiltered states $\rhoExp$ and the filtered states $\rho_{{\text{expt}},f}(v)$.}
	\label{Fig:rho_matrix}
\end{figure*}

From the definition of $\rhoWvd$ given in Eq.~(1) in main text and the explicit form of the swap operator:
\begin{equation} 
	V=\sum_{i,j} (\ket{i}\!\bra{j})_A\otimes(\ket{j}\!\bra{i})_B,
\end{equation}
we see that in the computational basis $\{\ket{i}\}_{i=0}^{d-1}$, Werner states $\rhoWvd$ may be represented by a density matrix using only {\em real} matrix elements, likewise for the filtered two-qubit density operator $\rhoWfd$, and the rotated version~\cite{LiPRR2021} prepared in our experiment:
\begin{align}\label{Eq:ActualFilteredState}
	\widetilde{W}^{(3)}_f(v)&=(\sigma_x\otimes\sigma_z)\rhoWf{3}(\sigma_x\otimes\sigma_z)^\dag,\\
	& = \frac{1}{N_{3,v}}\left[4(1-v)\proj{\Phi^+_2}+2v(\Id_4-\proj{\Phi^+_2})\right],\nonumber
\end{align}
where $N_{3,v}=4(1-v)+6v$ and $\ket{\Phi^+_2} = \frac{1}{\sqrt{2}}(\ket{1}\ket{2}+\ket{2}\ket{1})$. As an illustration of the quality of the experimentally prepared states, we show in \cref{Fig:rho_matrix} the real part of the unfiltered and filtered states prepared in the experiment.

It is also conventional to use a fidelity measure~\cite{LYM+19} to quantify the quality of the experimentally prepared state with respect to the target state. In particular, a widely adopted choice for quantifying the similarity between two density matrices $\rho$ and $\sigma$ is given by the Uhlmann-Jozsa fidelity:
\begin{equation}\label{Eq:Fidelity}
	F(\rho,\sigma)=\left[\tr\sqrt{ \sqrt{\rho} \sigma \sqrt{\rho} } \right]^2 = F(\sigma,\rho).
\end{equation}
Accordingly, we show in~\cref{tab:fidelity} the fidelity of the experimentally prepared two-qutrit states $\rhoExp$ and their filtered two-qubit counterparts $\rhoExpf$ with respect to, respectively, $\rhoWv{3}$ and $\widetilde{W}^{(3)}_f(v)$.

\begin{table}[htbp]
  \centering
  \caption{\label{Tbl:Fidelity} 
  Fidelity between the experimentally prepared states $\rhoExp$ [$\rhoExpf$] and their theory counterparts $\rhoWv{3}$ [$\widetilde{W}^{(3)}_f(v)$].}
    \begin{tabular}{|l|c|c|} 
    \hline
      $v$     & $F[\rhoExp,\rhoWvd]$& $F[\rhoExpf,\widetilde{W}^{(3)}_f(v)]$ \\ \hline 
      $0$     & $0.958 \pm 0.002$&  $0.934 \pm 0.006$ \\ \hline 
      $0.05$     & $0.965 \pm 0.001$& $0.948 \pm 0.001$\\ \hline 
      $0.1$     & $0.963 \pm 0.001$& $0.965 \pm 0.003$  \\ \hline 
      $0.15$     & $0.981 \pm 0.005$ & $0.974 \pm 0.003$ \\ \hline 
      $0.2$     & $0.982 \pm 0.004$& $0.985 \pm 0.002$\\ \hline 
      $0.25$     & $0.990 \pm 0.001$& $0.975 \pm 0.002$ \\ \hline 
      $0.3$     & $0.991 \pm 0.001$& $0.990 \pm 0.001$ \\ \hline 
      $0.35$     & $0.995 \pm 0.001$& $0.992 \pm 0.001$\\ \hline 
      $0.4$     & $0.980 \pm 0.003$& $0.995 \pm 0.001$  \\ \hline 
      $0.45$     & $0.993 \pm 0.001$& $0.995 \pm 0.001$ \\ \hline 
      $0.5$     & $0.994 \pm 0.001$& $0.994 \pm 0.001$ \\ \hline 
    \end{tabular}%
  \label{tab:fidelity}%
\end{table}%

\subsection{Steering robustness}
\label{App:ExpSR}

For completeness, we provide here, in comparison with \cref{fig:SR_W}, our best lower bounds on the steering robustness of the experimentally prepared states utilizing up to $n_s=7$ measurement bases. Notice that due to the asymmetry in the experimentally produced states, \cref{Eq:SR:AtoB} no longer coincides with \cref{Eq:SR}. Consequently, our results for SR$_{A\to B}$ shown in~\cref{fig:SR_W}, even though they still serve as a legitimate lower bound, may not represent the tightest lower bound on SR. Nonetheless, it is clear from \cref{fig:SR_W} that for any given $n_s$ and a non-vanishing interval of $v$, we see an improvement in these lower bounds on SR after the implemented ScLF.

\begin{figure}[t]
	\includegraphics[width=\columnwidth]{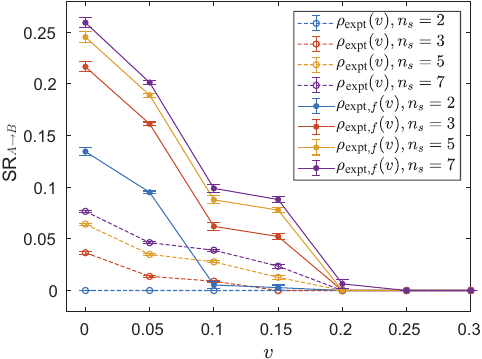}
	\caption{Best lower bounds on the steering robustness $\SR_{A\to B}$ found for $\rhoExp$ (hollow markers) and $\rhoExpf$ (filled markers) for various number of Alice's measurement settings $n_s$. The lines are guides for the eye.}
	\label{fig:SR_WExp}
\end{figure}

\subsection{Dense-codability}
\label{App:ExpDenseCodability}

\begin{figure}[h!]
	\includegraphics[scale=1]{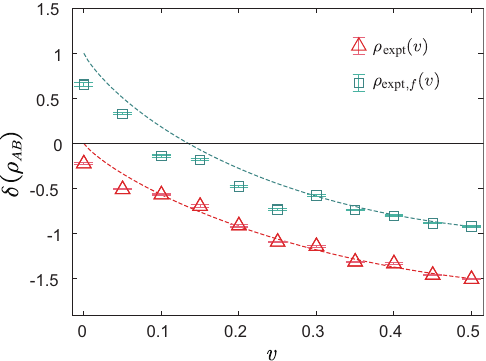}
	\caption{A bipartite quantum state $\rhoAB$ is useful for dense coding~\cite{Bruss2004densecoding} if the quantity $\delta(\rhoAB) := S(\rho_\text{B}) - S(\rhoAB) > 0$. Triangular (red) and square (green) markers show, respectively, $\delta[\rhoExp]$ and $\delta[\rhoExpf]$. Accordingly, the lower and upper dashed lines show $\delta[\rhoWvd]$ and $\delta[\rhoWfd]$.}
	\label{fig5:ds}
\end{figure}

Similarly, we provide in \cref{fig5:ds}, for completeness, a plot of the dense-codability of the experimentally prepared states before and after the ScLF. Again, the results clearly demonstrate the activation of dense codability for $v=0$ and $0.05$. However, the expected activation for $v=0.10$ is not observed in our imperfect experimental implementation.

%\bibliography{WernerStateRef.bib}
\bibliography{arXiv_v2.bbl}
\end{document}